\documentclass[a4paper,11pt]{article}
\usepackage{jcappub} 
\usepackage{lineno}

\usepackage{graphicx} 
\usepackage[dvipsnames]{xcolor}
\usepackage{rotating}
\usepackage{amssymb}
\usepackage{amsfonts}
\usepackage{amsmath}
\usepackage{xspace}
\usepackage{mathtools} 
\usepackage{verbatim}
\usepackage{comment}
\usepackage{enumitem}
\usepackage{physics}
\usepackage{graphicx}
\renewcommand{\t}{\eta}

\newcommand{\new}[1]{{\textcolor{black}{#1}}}

\begin{document}

\arxivnumber{2504.08062}
\title{From kinetic gases to an exponentially expanding universe - The Finsler-Friedmann equation}

\author[1,2]{Christian Pfeifer,}
\emailAdd{christian.pfeifer@zarm.uni-bremen.de}
\affiliation[1]{ZARM, University of Bremen, 28359 Bremen, Germany}

\author[2]{Nicoleta Voicu,}
\author[2]{Annamária Friedl-Szász}
\author[2]{and Elena Popovici-Popescu}
\affiliation[2]{Faculty of Mathematics and Computer Science, Transilvania University, Iuliu Maniu Str. 50, 500091 Brasov, Romania}

\abstract{
We investigate the gravitational field of a kinetic gas beyond its standard derivation from the second moment of the one-particle distribution function (1PDF), which typically serves as the energy-momentum tensor in the Einstein equations. This standard procedure raises an important question: why do the other moments of the 1PDF (which are needed to fully characterize the kinematical properties of the gas) not contribute to the gravitational field? Moreover, could these moments be relevant in addressing the dark energy problem? Using the canonical coupling of the full 1PDF to Finsler spacetime geometry via the Finsler gravity equation, we show that in general all moments contribute non-trivially. We derive the Finsler gravity equation in homogeneous and isotropic symmetry in conformal time -- dubbed the \emph{Finsler-Friedmann equation} -- which determines both the scale factor and the causal structure dynamically. Remarkably, this equation naturally admits a vacuum solution describing an exponentially expanding spacetime, without requiring a cosmological constant or any additional quantities. The resulting causal structure is a mild deformation of the one of Friedmann–Lemaître–Robertson–Walker (FLRW) geometry; close to the rest frame defined by cosmological time (i.e.\ for slowly moving objects), the causal structures of the two geometries are nearly indistinguishable.
}

\maketitle


\section{Introduction}
A large number of multi-particle systems, such as accretion discs \cite{Andreasson:2012cq,Rioseco:2016jwc,Gamboa:2021ytn,Cieslik:2022wok}, ordinary and neutron stars, gas clouds as well as the universe as a whole \cite{Astorga:2014cka,Ray1982}, can be described as kinetic gases \cite{Ehlers2011}: \new{a statistical mechanics description which is more fundamental compared to the often employed (perfect) fluid description \cite{Ames:2016coj,Andreasson:2021lsh}}. The properties of a kinetic gas are encoded in the so-called 1-particle distribution function $\varphi$ (1PDF), defined on the 1-particle phase space of spacetime, that is, on either the tangent or the cotangent bundle of spacetime \cite{Andreasson:2011ng,Sarbach:2013fya,Sarbach:2013uba,Acuna-Cardenas:2021nkj}. Its dynamics are described by the relativistic Boltzmann equation, a partial differential equation on phase space determining the behavior of the 1PDF along the trajectories of the constituents:
\begin{align}\label{eq:relBoltz}
    \dot x^a\frac{\partial}{\partial x^a} \varphi(x,\dot x) +\ddot x^a \frac{\partial}{\partial \dot x^a} \varphi(x,\dot x) = C\,.
\end{align}
The source term $C$ accounts for particle collisions and reactions. For a vanishing collision rate $C=0$, this equation is known as Liouville equation.

In contrast to this phase space treatment of the dynamics of a kinetic gas, its gravitational field is usually derived from the Einstein equations sourced by an energy-momentum tensor that lives on spacetime, obtained by an averaging procedure. More precisely, the energy-momentum tensor is defined as the second moment of the 1PDF with respect to the normalized particle 4-velocities,
\begin{align}\label{eq:KGEMtens}
    T^{ab}_{KG} = \int_{\mathcal{S}_x} d\Sigma_x \frac{\dot x^a \dot x^b}{g(\dot x, \dot x)} \varphi(x,\dot x)\,,
\end{align}
where $g$ is a spacetime metric, $\mathcal{S}_x$ is the set of normalized $4$-velocities and $d\Sigma_x$ is a suitable volume form \cite{Ehlers2011,Sarbach:2013uba,Sarbach:2013fya,Andreasson:2011ng,Acuna-Cardenas:2021nkj}. \new{Equations \eqref{eq:relBoltz} and \eqref{eq:KGEMtens} are known as the Einstein-Boltzmann, or,  in the particular case $C=0$, as the Einstein-Vlasov system. They combine general relativity and statistical mechanics in a consistent way.} However, from a phase space standpoint, the above gas-gravity coupling in \eqref{eq:KGEMtens} via averaging is rather specific -- and leads to two immediate questions: Why is only the second moment of the 1PDF the \new{direct} source of the gravitational field of the gas? How do the infinitely many other moments of the 1PDF contribute to the gravitational field?

The latter question can nicely be answered if one considers an extension of general relativity, based on Finsler geometry \cite{Rutz,Pfeifer:2011xi,Lammerzahl:2012kw,Lammerzahl:2018lhw,Pfeifer:2019wus,Villasenor:2022-1,Kapsabelis:2023khh}. Since the defining function of Finsler geometry lives on the 1-particle phase space of spacetime, it can naturally be coupled to the 1PDF of a kinetic gas, without resorting to any averaging \cite{Hohmann:2019sni}. From the canonical action-based Finsler gravity equation on phase space, effective field equations on spacetime can be obtained for \textit{any} moment of the 1PDF with respect to the velocities. Thus, Finsler gravity, which is a straightforward generalization of general relativity, offers the opportunity to understand the gravitational field of a kinetic gas from a new perspective. \new{It puts the gravitational field on the same statistical mechanics footing as the matter field, in contrast to the known approach which reduces the matter system (the kinetic gas) from its more fundamental statistical mechanics treatment to an effective description.} \vspace{10pt}

The goal is to study \new{if this new perspective on the gravitational field of kinetic gases}, in particular if the contributions of the higher moments of kinetic gases to their gravitational field, can be the source of the dark matter and dark energy phenomena \cite{Hohmann:2020yia}. In this article, we show that this is indeed the case, already for vacuum \new{solutions of the Finsler gravity equation -- which, in the non-vacuum case, couples the 1PDF of a kinetic gas directly to spacetime geometry}. 
\new{We find an exponentially expanding vacuum universe without the need of a cosmological constant, as a purely geometric effect.} The obtained geometry is indistinguishable from FLRW pseudo-Riemannian spacetime geometry for massive particles and observers propagating slowly with respect to the preferred cosmological rest frame; \new{deviations emerge at large velocities.}

\vspace{10pt}

\noindent This remarkable result is obtained \new{in Section \ref{sec:FriedSol}. The preceding sections introduce the necessary mathematical language and the general emergence of the Finsler-Friedmann equations from Finsler gravity in homogeneous and isotropic symmetry. The detailed outline of the article is as follows:}

In Section \ref{sec:FinsGRav}, we consider homogeneous and isotropic Finsler spacetimes of separated variables type, meaning that their time dependence can be completely absorbed into one conformal factor, interpreted as the scale factor of spacetime. This means we consider Finsler spacetimes on which there exists a conformal cosmological time. 

Using this ansatz in the Finsler gravity equation sourced by the 1PDF of a kinetic gas, we find the conformal time, canonical \emph{Finsler-Friedmann equation} in Section \ref{sec:Fried}. \new{The name is justified, as we will demonstrate, because in this case the Finsler gravity equation takes the form of a sum of three terms proportional to the spatial curvature of the universe $k$, the squared conformal Hubble function $\mathcal{H}^2$, and its conformal time derivative $\dot{\mathcal{H}}$—the same structure as the Friedmann equations derived from general relativity.} The coefficients of these terms are velocity-dependent Finsler geometric expressions. We prove that, by taking any moment of this equation with respect to the unit-normalized 4-velocities, one obtains $(0,n)$-tensor density equations on spacetime for any $n$. This is the answer to the question how all the moments of the 1PDF gravitate. All of them, and not just the second moment ($n=2$), are needed for a complete characterization of the gravitational field of a kinetic gas. \new{These $n$ equations show how not only the matter field degrees of freedom on spacetime can be determined from averaging, but also the gravitational (geometric) degrees freedom, thus putting matter and gravity on the same footing. Even better, the averaging can be done for interpretational reasons -- and is always possible by the integrability of the 1PDF -- but is not necessary in order to solve the equations. The Finsler gravity equation on the tangent bundle serves as master equation for all the effective equations on spacetime, meaning that solving the one scalar equation on the tangent bundle ensures that all resulting effective tensor densities equations are satisfied.}


In Section \ref{sec:FriedSol}, we discuss how the time evolution and the causal structure of spacetime are determined from the Finsler-Friedmann equation. For a non-vanishing 1PDF as source, the dynamics of these two properties of spacetime decouple partly, while for the vacuum equation, they decouple completely. For the latter case, we solve the conformal time dependence of the kinetic gas vacuum universe in full generality. Among the obtained solutions, when passing to cosmological time, we find an accelerated (exponentially) expanding universe. The corresponding causal structure (velocity dependence) of the corresponding Finsler Lagrangian is then determined by solving the remaining equation numerically. 

Finally, in Section \ref{sec:disc} we discuss the physical interpretation of this new perspective on the gravitational field of the kinetic gas in more detail. We conclude the article in Section \ref{sec:conc}. \vspace{10pt}

Overall, \new{our new approach, which we outlined here and discuss in any detail throughout this article,} leads to an improved, self-consistent coupling between physical systems described as kinetic gases and gravity, described as the Finslerian geometry of spacetime. Since we already find an exponentially expanding spacetime in vacuum, there is a promising prospect of explaining the observed early and late-time accelerated expansion of the universe using the non-vacuum Finsler gravity equation. \vspace{10pt}

Throughout this article we use the following notation: $M$ is a $4$-dimensional manifold and $TM$ is its ($8$-dimensional) tangent bundle. $(U, x^a)$ denotes a local chart on $M$, that leads to induced local coordinates $(x^a,\dot x^a)$ on $TU \subset TM$.  Latin indices $a,b,c$ run over the set $0,1,2,3$, while Greek indices $\alpha, \beta, \gamma$ can take the values $ 1,2,3$. Moreover $\partial_a  = \frac{\partial}{\partial x^a}$ and $\dot \partial_a  = \frac{\partial}{\partial \dot x^a}$ denote the canonical coordinate bases of the tangent spaces to the tangent bundle.

\section{Homogeneous and isotropic Finsler gravity}\label{sec:FinsGRav}
We start by briefly introducing the mathematical preliminaries we need, in order to discuss the Finsler-Friedmann equation, that describes spacetime dynamics, as well as the Boltzmann or Liouville equation which encodes the dynamics of the kinetic gas.

Finsler geometry is a straightforward generalization of Riemannian geometry, in which the geometry of a manifold is derived from a general spacetime line element $\sqrt{|L|}$ and its canonical Cartan nonlinear connection \cite{BCS,Bucataru}, instead of a spacetime metric $g$ and its Levi-Civita connection. In particular, Finsler \emph{spacetime} geometry has been developed in detail \cite{Beem, math-foundations,Bernal2020}, which paved the way to a Finslerian theory of gravity extending general relativity. \vspace{10pt}

A Finsler spacetime $(M,L)$ is a $4$-dimensional manifold $M$ equipped with a Finsler Lagrangian $L$, i.e., a smooth mapping $L:\mathcal{A}\to \mathbb{R}$ (where $\mathcal{A} \subset TM$ is a conic subbundle) with the following properties: 
\begin{itemize}
    \item          $L$ is positively $2$-homogeneous in the $\dot x$ variables, $L(x,\lambda \dot x)=\lambda^2 L(x,\dot x), \forall \lambda>0$ ;  
    \item $L$ possesses a nondegenerate \textit{Finsler metric} $g=g(x,\dot{x})$, given, in any chart of $\mathcal{A}$, by:
    \begin{align}
        g_{ab}(x,\dot x) = \frac{1}{2}\dot \partial_a \dot \partial_b L(x,\dot x)\,,
    \end{align}
    such that a well defined causal structure exists. 
    
    Technically, the latter means that there exists a conic subbundle $\mathcal{T}\subset \mathcal{A}$ with connected fibers $\mathcal{T}_x = \mathcal{T}\cap T_xM$,  $\forall x\in M$, such that on each $\mathcal{T}_x$ we have:  $L>0$, $g_{ab}$ has Lorentzian signature $(+,-,-,-)$ and $L$ can be continuously prolonged as 0 to the boundary $\partial\mathcal{T}_x$ of $\mathcal{T}_x$. 
    
    Physically, the set $\mathcal{T}$ is interpreted as the set of future pointing timelike directions.
\end{itemize}

Besides $\mathcal{T}$, this definition of Finsler spacetime ensures the existence of the following additional relevant sets of directions:
\begin{itemize}
    \item The conic set of admissible vectors $\mathcal{A}\subset TM$, where $L$ is smooth and all Finsler geometric objects are well defined.
    \item A cone of non-trivial null directions $\mathcal{N} = \{(x,\dot x)\in TM | L(x,\dot x) = 0\}$, describing the propagation of light.
    \item The set $\mathcal{S} = \bigcup_{x\in M}\mathcal{S}_x$, with $\mathcal{S}_x :=\{(x,\dot x)\in \mathcal{T}_x | L(x,\dot x) = 1\}$, of all normalized future pointing timelike directions, identified as physical $4$-velocities.
\end{itemize}
These technical requirements ensure that we can consistently study physical systems and their evolution on a Finsler spacetime. For all details on the definition of  a Finsler spacetime we refer to~\cite{Beem, math-foundations,Bernal2020}.\vspace{10pt}

The canonical action based Finsler gravity equations are obtained as Euler-Lagrange equations from the canonical generalization of the Einstein-Hilbert action to Finsler spacetimes \cite{Pfeifer:2011xi,Hohmann_2019,Hohmann:2019sni}, as
\begin{align}\label{eq:fgrav}
    3\frac{ R}{L} -\frac{1}{2}g^{ab}\dot{\partial}_{a}\dot{\partial}_{b}R  + g^{ab}\left(\nabla_{\delta_{a}}P_{b}-P_{a}P_{b}+\dot{\partial}_{a}(\nabla P_{b})\right) = \kappa \varphi \,.
\end{align}
Here, $\kappa$ is the Finsler gravitational constant (to be determined from the appropriate general relativity or Newtonian limit), $\varphi$ is the 1PDF of the kinetic gas which acts as a source of the gravitational field, $R$ is the Finsler-Ricci scalar, $P$ is the Landsberg tensor, $\nabla = \dot x^a \nabla_{\delta_{a}}$ is the dynamical covariant derivative and $\nabla_{\delta_{a}}$ the Chern-Rund-linear covariant derivative on $TM$, the definition of these objects can be found in Appendix~\ref{app:FinsGeom}.

The Finsler gravity equation  can compactly be written as 
\begin{align}\label{eq:fgrav2}
    \mathcal{G}(x,\dot x) = \kappa \varphi(x,\dot x)\,.
\end{align}
It shows how all the information contained in the 1PDF contributes to the gravitational field of the kinetic gas, described by the Finslerian geometry of spacetime. Applying an integral over the set of all unit timelike directions $\mathcal{S}_x$ at $x \in M$, generates effective gravitational field equations on spacetime for each moment of the 1PDF:
\begin{align} \label{eq:moments_general}
    &G^{a_1...a_n}(x) =\\
    & \int_{\mathcal{S}_x} \frac{\dot x^{a_1} ...\dot x^{a_n}}{L^{\frac{n}{2}}}\ \mathcal{G}(x,\dot x) \ d\Sigma_x\
    = \kappa \int_{\mathcal{S}_x} \frac{\dot x^{a_1} ...\dot x^{a_n}}{L^{\frac{n}{2}}} \ \varphi(x,\dot x)\ d\Sigma_x\\\
    &=\kappa T^{a_1...a_n}(x)\,.
\end{align}
For simplicity, it is often assumed that $\varphi$ is compactly supported on the set of unit timelike directions \cite{Sarbach:2013uba,Acuna-Cardenas:2021nkj}, at least it must be integrable, since the zeroth moment of the 1PDF gives the particle number density on spacetime, which is finite.

For $n=2$, the equations resemble the form of the Einstein equations
\begin{align}
    G^{ab}(x) = \kappa T^{ab}(x)\,.
\end{align}
Despite this formal resemblance, the left-hand side of \eqref{eq:moments_general} is generally expected to differ from the Einstein tensor, in the sense that it acquires additional and different terms through the integration procedure. It is still under investigation under which conditions the Einstein tensor is reproduced. One conjecture is that it might emerge when the integral is only taken around unit timelike directions that describe massive objects with small velocities or at rest. 

The gravitational field equations are complemented by the Finsler generalization of the relativistic Boltzmann equation \eqref{eq:relBoltz}, which takes the form
\begin{align}\label{eq:FinBolz}
    \nabla \varphi = C\,;
\end{align}
this becomes the so-called Finsler Liouville equation for $C=0$. The scalar $C$ is called the collision rate and its vanishing means that the number of particles is conserved, as proven in detail in \cite{HPV2020}. 
Equations \eqref{eq:fgrav} and \eqref{eq:FinBolz} are the Finsler extension of the Einstein-Vlasov equations, which treat gravity and matter on the same footing. They take all properties of the kinetic gas \new{directly} into account, when deriving its gravitational field, and do not ignore any moments. Physically speaking, the Finsler gravity equation determines not only the evolution and the position dependence of the gravitational field on spacetime (encoded in the $x$-dependence of the geometry defining Finsler Lagrangian $L$), but also the causal structure of spacetime such as light cones and unit timelike directions (encoded in the $\dot x$-dependence of the geometry defining Finsler Lagrangian $L$). \vspace{10pt}

To demonstrate explicitly how to find physically viable solutions to the Finsler gravity equations and how averaging over unit timelike velocities leads to effective field equations on spacetime, we study the equations in cosmological, homogeneous and isotropic symmetry. Homogeneous and isotropic Finsler spacetime functions $L$ are defined in spherical local coordinates $(x^a, \dot x^a) = (t,r,\theta,\phi,\dot t, \dot r, \dot \theta, \dot \phi)$ by, see \cite{Hohmann:2020mgs},
\begin{align}\label{eq:homisoL}
	L(t,r,\theta,\phi,\dot t, \dot r, \dot \theta, \dot \phi) = L(t,\dot t, w)\,,\quad w^2 = \frac{\dot r^2}{1-k r^2}+r^2\big(\dot \theta^2+\sin^2(\theta) \dot\phi^2\big)\,,
\end{align}
where $k$ is the spatial curvature parameter of the universe. Due to the homogeneity property of $L$, it can conveniently be expressed in terms of a new variable $s=w/\dot t$\new{, which can be interpreted as spatial velocity with respect to the cosmological rest frame ($w=0$),} as
\begin{align}\label{eq:homIsoL}
    L(t,\dot t, w) = \dot t^2 L(t,1, s) =: \dot t^2 h(t,s)^2\,.
\end{align}
In general, it is unclear how to display the Finsler gravity equation \eqref{eq:fgrav} for general homogeneous and isotropic Finsler Lagrangians $L$ in a readable and compact way, as one can see from the Finsler geometric objects displayed in Appendix \ref{app:HomAndIsoGeom}. Therefore, we continue with a very promising subclass \new{that is defined by the existence of conformal time.}

\section{The Finsler-Friedmann Equation}\label{sec:Fried}
We turn our discussion to the dynamical equations of homogeneous and isotropic Finsler gravity. We first discuss the gravitational field equations before we turn to the Boltzmann and Liouville equations at the end of this section. 

The function $h(t,s)$ defining the homogeneous and isotropic Finsler Lagrangian \eqref{eq:homIsoL} encodes, from a spacetime perspective, in principle infinitely many degrees of freedom; this is seen as, for each value of $s$, there is one free function of $t$. Only when the $s$-dependence is fixed by some principle or through some equation, the number of $t$-dependent functions will be specified. This is a vast generalization, compared to the situation of Friedmann-Lemaitre-Robertson-Walker (FLRW) spacetime geometry, where only the scale factor $a(t)$ and the value of the spatial curvature $k$ are present as degrees of freedom, whereas the $s$-dependence is \emph{a priori} fixed. \vspace{10pt}

In order to find Finsler Lagrangians close to pseudo-Riemannian geometry, we consider a conformal time, separated variable ansatz of the type
\begin{align}\label{eq:LHomIso}
    L(\t, s) = \dot \t^2 a(\t)^2 f(s)^2\,,
\end{align}
for which we renamed the general time coordinate $t$ to $\t$, as it resembles conformal time in FLRW geometry\footnote{\new{This class in particular also contains the homogeneous and isotropic Landsberg spacetimes, i.e.\ those Finsler spacetimes for which the Landsberg tensor $P$ is identically zero \cite{Friedl-Szasz:2024vtu} and the Finsler gravity equation simplifies.}}. The conformal time scale factor $a(\t)$ encodes the spacetime evolution of the gravitational field, while the function $f(s)$ determines the causal structure of spacetime (in other words, the 4-velocity dependence of the Finsler Lagrangian). We note that the canonical, unique Landsberg-type generalization of FLRW spacetime geometry is of this form and that the explicit expression of $f(s)$ for this case has been derived in~\cite{Friedl-Szasz:2024vtu}. In this language, pseudo-Riemannian FLRW spacetime geometry in conformal time is given by choosing $f(s) = \sqrt{1-s^2}$, since then 
\begin{align}
    L &= \dot \t^2 a(\t)^2 (1 - s^2) 
    = a(\t)^2 (\dot \t^2 - w^2)\\
    &= a(\t)^2 \left(\dot \t^2 - \frac{\dot r^2}{1-kr^2} - r^2\dot \theta^2 - r^2 \sin^2(\theta) \dot\phi^2\right)\,.
\end{align}
From the phase space perspective that we take in this article, $f(s)$ is not chosen \emph{a priori} but determined from dynamical equations.\vspace{10pt}

For a general Finsler Lagrangian of the form \eqref{eq:LHomIso}, the Finsler gravity equation \eqref{eq:fgrav} nicely splits into terms that also have separated variables. We will call the obtained equation, the \emph{Finsler-Friedmann equation}
\begin{align}\label{eq:FinsFried} 
\frac{k}{a(\eta)^2}\ \mathcal{G}_k(s) + \frac{\mathcal{H}^2(\t)}{a(\eta)^2}\ \mathcal{G}_\mathcal{H}(s) + \frac{\dot {\mathcal{H}}(\eta)}{a(\eta)^2}\ \mathcal{G}_{\dot {\mathcal{H}}}(s) = \kappa\ \varphi(\t,s)\,,
\end{align}
where $\mathcal{H}(\t) = \frac{\dot a(\t)}{a(\t)}$ (here "dot" means derivative w.r.t. $\t$), and the functions $\mathcal{G}_k$, $\mathcal{G}_\mathcal{H}$ and $\mathcal{G}_{\dot {\mathcal{H}}}$ are built from $f$ and its derivatives. Their explicit form, shown in Appendix \ref{app:G}, is derived from the objects displayed in Appendix \ref{app:HomAndIsoGeom}. In homogeneous and isotropic symmetry, the 1PDF can only depend on $\t$ and $s$.

In \eqref{eq:FinsFried}, we clearly see the form of a Friedmann-type equation emerging, where the functions $\mathcal{G}_k,$ $\mathcal{G}_\mathcal{H}$ and $\mathcal{G}_{\dot {\mathcal{H}}}$ are the prefactors of the spatial curvature term $k$, the Hubble function squared $\mathcal{H}^2$ and, respectively, its derivative $\dot {\mathcal{H}}$. Assuming that the matter content of the universe is regarded as a kinetic gas, described by the 1PDF $\varphi$, this equation determines the gravitational field without losing any information about the gas and its imprint on spacetime geometry -- and thus, on gravity.

Due to the specific form of equation \eqref{eq:FinsFried}, we find the following necessary conditions, which partly separate the time evolution and the causal structure (direction dependence) of the geometry 
\begin{align}\label{eq:splitting-eta-s}
   0 = \frac{d}{d\t}\left(\frac{\frac{d}{d\t}\left(\frac{1}{2 \mathcal{H} \dot {\mathcal{H}}}\frac{d}{d\t} \left(\kappa\ a^2 \varphi(\t,s)\right)\right)}{\frac{d}{d\t}\left(\frac{\ddot {\mathcal{H}}}{2 \mathcal{H} \dot {\mathcal{H}} }\right)}\right)\,,\quad\quad
   0 = \frac{d}{ds}\left( \frac{\frac{d}{ds} \left( \frac{ \frac{d}{ds}\left( \frac{\varphi(\eta, s)}{\mathcal{G}_k}\right) }{ \frac{d}{ds}\left( \frac{\mathcal{G}_\mathcal{H}}{\mathcal{G}_k}\right) } \right) }{ \frac{d}{ds} \left( \frac{\frac{d}{ds}\left( \frac{\mathcal{G}_{\dot{\mathcal{H}}}}{\mathcal{G}_k} \right) }{ \frac{d}{ds} \left( \frac{\mathcal{G}_\mathcal{H}}{\mathcal{G}_k} \right) } \right) } \right)\,.
\end{align}

Another aspect of \eqref{eq:FinsFried} is that one can obtain infinitely many \emph{Finsler-Friedmann tensor-density equations} on spacetime by considering the $n$-th moment of the original field equation with respect to the 4-velocities:
\begin{align}\label{eq:FinsFriedInt}
    &\tfrac{k}{a^2} \int_{\mathcal{S}_x} \left(\tfrac{\dot x^{a_1}...\dot x^{a_n}}{L^{\frac{n}{2}}} \mathcal{G}_k(s) \dd \Sigma_x\right)|_{\mathcal{S}_x}
    + \tfrac{\mathcal{H}^2}{a^2} \int_{\mathcal{S}_x} \left(\tfrac{\dot x^{a_1}...\dot x^{a_n}}{L^{\frac{n}{2}}} \mathcal{G}_\mathcal{H}(s) \dd \Sigma_x\right) |_{\mathcal{S}_x}
    + \tfrac{\dot{\mathcal{H}}}{a^2} \int_{\mathcal{S}_x} \left(\tfrac{\dot x^{a_1}...\dot x^{a_n}}{L^{\frac{n}{2}}} \mathcal{G}_{\dot {\mathcal{H}}}(s)  \dd \Sigma_x\right)|_{\mathcal{S}_x}\nonumber\\
    &= \kappa \int_{\mathcal{S}_x} \left(\tfrac{\dot x^{a_1}...\dot x^{a_n}}{L^{\frac{n}{2}}}  \varphi(\t,s) \dd \Sigma_x\right)|_{\mathcal{S}_x}\,,
\end{align}
here, $\mathcal{S}_x$ is the set of unit-normalized future pointing $4$-velocities introduced in Section \ref{sec:FinsGRav} and $\dd \Sigma_x$ is the canonical Finslerian volume form on  $\mathcal{S}_x$, expressed as
\begin{align}
    \dd \Sigma_x =  \frac{-\det g}{L^2} i_\mathbb{C}(d^4x\wedge d^4\dot x)\,,
    \quad \mathbb{C} = \dot x^a \dot \partial_a\,.
\end{align}
Thus, from the pure spacetime point of view, the integrated field equations \eqref{eq:FinsFriedInt} can be expressed as tensor density equations
\begin{align}\label{eq:FriedTensDen}
    \frac{k}{a^2} G_k^{a_1...a_n} + \frac{\mathcal{H}^2}{a^2}G_{{\mathcal{H}}}^{a_1...a_n} + \frac{\dot{\mathcal{H}}}{a^2} G_{\dot {\mathcal{H}}}^{a_1...a_n} = \kappa T^{a_1...a_n}\,,
\end{align}
where the coefficient tensor-densities can symbolically be written as
\begin{align}
    G_k^{a_1...a_n} &= \int_{\mathcal{S}_x} \left(\tfrac{\dot x^{a_1}...\dot x^{a_n}}{L^{\frac{n}{2}}} \mathcal{G}_k(s) \dd \Sigma_x\right)|_{\mathcal{S}_x}\,,\\
    G_{{\mathcal{H}}}^{a_1...a_n} &= \int_{\mathcal{S}_x} \left(\tfrac{\dot x^{a_1}...\dot x^{a_n}}{L^{\frac{n}{2}}} \mathcal{G}_\mathcal{H}(s) \dd \Sigma_x\right) |_{\mathcal{S}_x}\,, \\
    G_{\dot {\mathcal{H}}}^{a_1...a_n}&= \int_{\mathcal{S}_x} \left(\tfrac{\dot x^{a_1}...\dot x^{a_n}}{L^{\frac{n}{2}}} \mathcal{G}_{\dot {\mathcal{H}}}(s)  \dd \Sigma_x\right)|_{\mathcal{S}_x}\,.
\end{align}  
In general, one cannot guarantee that the individual integrals are finite. However, the integrability of the 1PDF ensures that their weighted sum, as displayed in \eqref{eq:FriedTensDen}, is always finite. In Appendix \ref{app:IntInd}, we show the details of how such integrals can be calculated explicitly. 

A consistent solution of equations \eqref{eq:FriedTensDen} for every $n$, which is equivalent to a solution of the original non-averaged field equation \eqref{eq:FinsFried}, gives the complete gravitational field of a kinetic gas. \vspace{10pt}

Let us have a closer look at the second moment equation, since it resembles the Einstein equations in a bit more detail,
\begin{align}\label{eq:FriedTensDenSecond}
    \frac{k}{a^2} G_k^{ab} + \frac{\mathcal{H}^2}{a^2} G_{{\mathcal{H}}}^{ab} + \frac{\dot{\mathcal{H}}}{a^2} G_{\dot {\mathcal{H}}}^{ab} = \kappa T^{ab}\,.
\end{align}
Just considering this equation alone, certainly does not capture all the aspects of the system. In particular, it fails to take into account all properties of the matter field encoded in the 1PDF $\varphi$ (from the phase space point of view), respectively, in the matter tensor densities $T^{a_1...a_n}$ (from the spacetime point of view).

In conformal time, the Einstein equations for a FLRW spacetime metric $\tilde g$ coupled to a perfect fluid, characterized through its energy density $\rho$ and its pressure $p$, take the form \eqref{eq:FriedTensDenSecond}, with
\begin{align}
    (G_k^{ab}) = (G_\mathcal{H}^{ab}) = 
    \begin{pmatrix}
    -3 & 0 & 0 & 0\\
    0 & -(1-k r^2) & 0 & 0\\
    0 & 0 & -\frac{1}{r^2} & 0\\
    0 & 0 & 0 & -\frac{1}{r^2 \sin^2\theta}
    \end{pmatrix}\,,
    \qquad
    (G_{\dot {\mathcal{H}}}^{ab}) =
    \begin{pmatrix}
    0 & 0 & 0 & 0\\
    0 & -2(1-k r^2) & 0 & 0\\
    0 & 0 & -\frac{2}{r^2} & 0\\
    0 & 0 & 0 & -\frac{2}{r^2 \sin^2\theta}
    \end{pmatrix}\,,
\end{align}
and
\begin{align}
    (T_{KG}^{ab}) = 
    \begin{pmatrix}
    \frac{\rho}{a^2} & 0 & 0 & 0\\
    0 & \frac{(1-k r^2) p}{a^2} & 0 & 0\\
    0 & 0 & \frac{p}{r^2 a^2} & 0\\
    0 & 0 & 0 & \frac{p}{r^2 \sin^2\theta}\,.
    \end{pmatrix}\,.
\end{align}
In contrast to the equations \eqref{eq:FinsFriedInt} which we derived from a phase space perspective, only the right-hand side of \eqref{eq:FriedTensDenSecond}, i.e., the energy-momentum tensor, is obtained through averaging \cite{Ehlers2011,Andreasson:2011ng,Sarbach:2013fya,Sarbach:2013uba,Acuna-Cardenas:2021nkj}, as
\begin{align}
    T^{ab}_{KG} = \int_{\tilde{\mathcal{S}}_x} \dd \tilde\Sigma_x \frac{\dot x^a \dot x^b}{\tilde g(\dot x, \dot x)} \varphi(x,\dot x)\,,
\end{align}
where $\dd \tilde\Sigma_x$ is the canonical volume form on the set of unit timelike future pointing directions $\tilde{\mathcal{S}}_x$ defined by the spacetime metric $\tilde g$.\vspace{10pt}

The relativistic Boltzmann equation for a Finsler function of the type \eqref{eq:LHomIso} becomes
\begin{align}\label{eq:Boltz}
    \nabla \varphi(\t,s)|_{\mathcal{S}} = \frac{1}{a(\t)f(s)} \left(\frac{\partial}{\partial \t} \varphi - \mathcal{H}(\t) \frac{f'}{f''} \frac{\partial}{\partial s} \varphi(\t,s)\right) = C\,,
\end{align}
which can easily be seen by setting $h(t,s)\to a(\t) f(s)$ in equation \eqref{eq:nablah}, and using that on $\mathcal{S}$ we have $\dot \t = 1/(a(\t)f(s))$. In particular, for a collisionless gas ($C=0$) \eqref{eq:Boltz} becomes the Liouville equation. Its integration leads to a 1PDF of the form:
\begin{align}
    \varphi(\t,s) = \varphi(a(\t) f'(s))\,.
\end{align}
These findings generalize the results presented in \cite{Astorga:2014cka,Acuna-Cardenas:2021nkj} for kinetic gases on FLRW spacetime.\vspace{10pt}

The phase space coupling between the matter degrees of freedom of the kinetic gas and the geometric gravitational degrees of freedom considers both of them on the same footing, in a very natural way.\vspace{10pt}

Having clarified the conceptual aspects of the Finsler-Friedmann equation, we turn towards finding solutions and to their physical consequences.

\section{The exponentially expanding vacuum solution}\label{sec:FriedSol}

On the basis of general relativity and of the Einstein equations, the only way to obtain a non-trivial \new{(non-Minkowski)} homogeneous and isotropic vacuum \new{spacetime} solution is to introduce a cosmological constant, which serves as the effective source of the vacuum dynamics. Usually, it is interpreted as one possible source of dark energy. 

Here, we find that the \new{vacuum solution of the Finsler gravity equation, which is capable of coupling the 1PDF of a kinetic gas directly and without averaging to the geometry of spacetime,} 
naturally allows for an accelerated expansion from its dynamical equation, without the need of introducing any cosmological constant or any further quantities. Thus, we can interpret the new Finsler gravity tangent bundle degrees of freedom of the gravitational field as a potential source of dark energy. This perspective is not available from the pure spacetime point of view.

Before we derive the exponentially expanding Finsler gravity kinetic gas universe in Section \ref{ssec:vacsol}, we briefly discuss, for completeness and future investigation, how the zeroth and first moments of the Finsler-Friedmann equation can be related to the standard form of the Friedmann equations on spacetime.

\subsection{The averaged equations on spacetime}\label{ssec:matsol}
Due to homogeneous and isotropic symmetry, we find two independent equations on spacetime from the zeroth and the first moment of the Finsler gravity equation. The derivation works as outlined in \eqref{eq:FriedTensDen} (and using the techniques from \ref{app:IntInd}):
\begin{itemize}
    \item The \emph{number density} equation, see also \ref{sapp:scalint},
    \begin{align}\label{eq:NumbDense}
        \kappa \hat N(x) 
        &= \kappa \int_{\mathcal{S}_x} \left(\varphi_L(\t,s) \dd\Sigma_x\right)|_{\mathcal{S}_x}\\
        &= \frac{k}{a^2} G_k + \frac{\mathcal{H}^2}{a^2}G_{{\mathcal{H}}} + \frac{\dot{\mathcal{H}}}{a^2} G_{\dot {\mathcal{H}}}\,,
    \end{align}    
    \item The only non-vanishing component \emph{particle-current density} equation, see also \ref{sapp:vecint},
    \begin{align}\label{eq:ParticleCurrentDense}
        \kappa \hat J^0(x)
        &= \kappa \int_{\mathcal{S}_x} \left( \dot \eta \varphi_L(\t,s)\dd\Sigma_x\right)|_{\mathcal{S}_x}\\
        &= \frac{k}{a^2} G_k^0 + \frac{\mathcal{H}^2}{a^2}G_{{\mathcal{H}}}^0 + \frac{\dot{\mathcal{H}}}{a^2} G_{\dot {\mathcal{H}}}^0\,.
    \end{align}
\end{itemize}
This system of equations can be solved for 
\begin{align}
    \frac{\mathcal{H}^2}{a^2}  
    &= \frac{k}{a^2} \left(\frac{G_k G_{\dot {\mathcal{H}}}^0 - G_k^0 G_{\dot {\mathcal{H}}} }{G_{\dot {\mathcal{H}}} G_{{\mathcal{H}}}^0 - G_{\dot {\mathcal{H}}}^0 G_{{\mathcal{H}}}}\right) 
    + \kappa \left(\frac{   G_{\dot {\mathcal{H}}} \hat J^0 - G_{\dot {\mathcal{H}}}^0 \hat N  }{G_{\dot {\mathcal{H}}} G_{{\mathcal{H}}}^0 - G_{\dot {\mathcal{H}}}^0 G_{{\mathcal{H}}}} \right) \,,\\
    \frac{\dot{\mathcal{H}}}{a^2}  
    &= \frac{k}{a^2} \left(\frac{G_{{\mathcal{H}}} G_{k}^0 - G_{{\mathcal{H}}}^0 G_{k} }{G_{\dot {\mathcal{H}}} G_{{\mathcal{H}}}^0 - G_{\dot {\mathcal{H}}}^0 G_{{\mathcal{H}}}} \right) + \kappa \left(\frac{ G_{{\mathcal{H}}}^0 \hat N - G_{{\mathcal{H}}} \hat J^0  }{G_{\dot {\mathcal{H}}} G_{{\mathcal{H}}}^0 - G_{\dot {\mathcal{H}}}^0 G_{{\mathcal{H}}}} \right) \,,
\end{align}
which are the counterparts to the usual Friedmann equations in general relativity, which, in conformal time,  take the form
\begin{align}
    \frac{\mathcal{H}^2}{a^2}  &= - \frac{k}{a^2} + \frac{\kappa}{3} \rho\,,\\
    \frac{\dot{\mathcal{H}}}{a^2} &= -\frac{\kappa}{6} (3 p+ \rho)\,.
\end{align}
All further non-vanishing moment equations that can be obtained from \eqref{eq:FriedTensDen} determine the tensor components $G_I^{a_1...a_n},\ I=k, \mathcal{H}, \dot {\mathcal{H}}$ from the higher moments of the 1PDF of the kinetic gas, and thus the function $f(s)$ in \eqref{eq:LHomIso}, which encodes the causal structure of spacetime.

\subsection{Vacuum solutions}\label{ssec:vacsol}
To demonstrate the impact of the change of perspective in the understanding of the gravitational field of a kinetic gas, we solve the Finsler-Friedmann equation in vacuum. In this case, it simplifies to
\begin{align}\label{eq:FinsFriedVac}
    k \mathcal{G}_k(s) + \mathcal{H}^2(\t)\ \mathcal{G}_\mathcal{H}(s) + \dot {\mathcal{H}}(\eta)\ \mathcal{G}_{\dot {\mathcal{H}}}(s) = 0\,,
\end{align}
and determines the scale factor $a$ as well as the direction dependence of the Finsler Lagrangian encoded in $f(s)$. In vacuum, we can decouple the $\eta$ and $s$ dependencies as follows. 

Taking a first $\eta$ derivative yields
\begin{align}
    2 \mathcal{H} \dot {\mathcal{H}} \ \mathcal{G}_\mathcal{H}(s) + \ddot {\mathcal{H}}\ \mathcal{G}_{\dot {\mathcal{H}}}(s) = 0
    \Rightarrow 2 \mathcal{G}_\mathcal{H}(s) = - \frac{\ddot {\mathcal{H}}\ \mathcal{G}_{\dot {\mathcal{H}}}(s)}{\mathcal{H} \dot {\mathcal{H}} }
\end{align}
which,  via another $\eta$ derivative then implies
\begin{align}\label{eq:dH=}
   \frac{d}{d\t} \frac{\ddot {\mathcal{H}}}{{\mathcal{H}} \dot {\mathcal{H}} } = 0 \Leftrightarrow \dot {\mathcal{H}}  = c_1 {\mathcal{H}}^2 + c_2\,.
\end{align}
Using this expression in \eqref{eq:FinsFriedVac}, we find the equivalent conditions
\begin{align}
   k \mathcal{G}_k(s) + c_2 \mathcal{G}_{\dot {\mathcal{H}}}(s) &= 0 \label{eq:s-dep1}\\
   \mathcal{G}_\mathcal{H}(s) + c_1 \mathcal{G}_{\dot {\mathcal{H}}}(s) &= 0 \label{eq:s-dep2}\,.
\end{align}

Equation \eqref{eq:dH=} completely determines the time evolution of the vacuum state of the homogeneous and isotropic \new{vacuum} universe, while \eqref{eq:s-dep1} and \eqref{eq:s-dep2} specify its causal structure by determining $f$, i.e., the velocity dependence of the geometry defining a Finsler Lagrangian. 

In the following, we identify self-consistent physical solutions of these equations. Most remarkably, we find an accelerated expanding universe.

\subsubsection{The time evolution of the vacuum state of the kinetic gas universe}
In order to find the evolution of the kinetic gas vacuum universe in cosmological time, we use the relation $a(\eta)d\eta=dt$, to express the conformal time Hubble function
\begin{align}
    \mathcal{H}(\eta(t)) = \frac{\dot a(\eta(t))}{a(\eta(t))} = \frac{da(\eta(t))}{dt}\,,
\end{align}
and the time-evolution equation \eqref{eq:dH=} as
\begin{align}
    a \frac{d^2 a}{dt^2} = c_1 \left(\frac{d a}{dt} \right)^2 + c_2\,.
\end{align}
Finding a closed general solution of this equation, using elementary functions, is not possible. However, it is possible to find solutions for specific values of $c_1$ and $c_2$. Among them, we find extraordinarily interesting cases when we choose $c_2 = 0$. There exist two branches for this case:

\begin{itemize}
\item For $c_1=1$, the cosmological time scale factor becomes:
\begin{align}\label{eq:expuni}
    a(t)=d_2 e^{d_1t}\,,\quad d_1, d_2 \in\mathbb{R}\,,
\end{align}
which is a most remarkable result. The Finsler-Friedmann equation in vacuum naturally leads to an exponentially expanding universe. \new{This is a first direct indication that the treatment of the gravitational field on the same footing as a kinetic gas, i.e., from a statistical physics perspective, could explain, at least in part, the dark energy content of the universe, whose origin would be those degrees of freedom which do not appear in the usual Einstein-Vlasov approach. We expect that this result percolates to the non-vacuum case that includes a non-trivial 1PDF as source.}


\item If $c_1\neq 1$, then,
\begin{align}\label{eq:c2=0-2}
a(t)= \left(d_1t+d_2\right)^{\frac{1}{1-c_1}}\,,\quad d_1, d_2 \in\mathbb{R}\,,
\end{align}
which reveals that the kinetic gas vacuum offers a variety of interesting dynamics for the universe.
\end{itemize}

Although we could not present a closed, general solution for arbitrary $c_1$ and $c_2$ of equation \eqref{eq:dH=} in cosmological time, we find the scale factor $a(\eta)$ in conformal time in full generality. The different branches that emerge are:
\begin{itemize}
    \item For $c_2 >0$:
    \begin{center}
    \begin{tabular}{|c| c| c| c|} 
    \hline
    $c_1$ & $>0$ & $=0$ & $<0$ \\
    \hline
    $\mathcal{H}(\t)$ 
    & $  \tfrac{\sqrt{c_2} \tan \left(\sqrt{c_1} \sqrt{c_2} (\t+c_3)\right)}{\sqrt{c_1}}$ 
    & $c_3 + c_2 \t$ 
    & $\tfrac{\sqrt{c_2} \tanh \left(\sqrt{|c_1|} \sqrt{c_2} (\t+c_3)\right)}{\sqrt{|c_1|}}$\\
    \hline
    $a(\t)$ 
    & $\tfrac{c_4}{|\cos\left(\sqrt{c_1} \sqrt{c_2} (\t+c_3)\right)|^{1/c_1}}$ 
    & $c_4 e^{\frac{c_2 \t^2}{2}+c_3 \t}$
    & $\tfrac{c_4}{(-1)^{1/(2 c_1)}\cosh\left(\sqrt{|c_1|} \sqrt{c_2} (\t+c_3)\right)^{1/c_1}}$ \\
    \hline
    \end{tabular}
    \end{center}
    \item For $c_2 = 0$, we do not need to conduct a case-by-case study for different choices of $c_1$, as we needed to do in cosmological time (where we found the solutions \eqref{eq:expuni} and \eqref{eq:c2=0-2}); in conformal time, the general solution is:
    \begin{align}\label{eq:c2=0}
        \mathcal{H}(\t) = -\frac{1}{c_1 \t + c_3}\,,\quad a(\t) = \pm \frac{c_4^2}{(c_3+c_1 \t)^{1/c_1}},.
    \end{align}
    \item For $c_2< 0$:
    \begin{center}
    \begin{tabular}{|c|c|c|c|} 
    \hline
    $c_1$ & $>0$ & $=0$ & $<0$ \\
    \hline
    $\mathcal{H}(\t)$ 
    & $- \tfrac{\sqrt{|c_2|} \tanh \left(\sqrt{c_1} \sqrt{|c_2|} (\t+c_3)\right)}{\sqrt{c_1}}$ 
    & $c_3 + c_2 \t$  
    & $- \frac{\sqrt{|c_2|} \tan \left(\sqrt{|c_1|} \sqrt{|c_2|} (\t+c_3)\right)}{\sqrt{|c_1|}}$ \\
    \hline 
    $a(\t)$ 
    & $\tfrac{c_4}{(-1)^{1/(2 c_1)}\cosh\left(\sqrt{c_1} \sqrt{|c_2|} (\t+c_3)\right)^{1/c_1}}$  
    & $c_4 e^{\frac{c_2 \t^2}{2}+c_3 \t}$  
    & $  \frac{c_4}{|\cos\left(\sqrt{|c_1|} \sqrt{|c_2|} (\t+c_3)\right)|^{1/c_1}}$ \\ 
    \hline
    \end{tabular}
\end{center}
\end{itemize}
\new{This list classifies all possible evolutions of a vacuum Finsler universe. Of course, the question of whether all of these correspond to solutions of \eqref{eq:s-dep1} and \eqref{eq:s-dep2} that represent physically viable Finsler spacetimes consistent with observations is a separate issue, which will be investigated in the future. In this article, we will limit ourselves to further studying the intriguing case of the exponentially expanding universe.}

\new{Before we solve Equations \eqref{eq:s-dep1} and \eqref{eq:s-dep2} numerically to answer the remaining open question: "What is the corresponding spacetime causal structure encoded in the remaining free function $f(s)$?", we like to get a better understanding of the origin of the exponential expanding vacuum solution of the Finsler gravity equation.}

\subsubsection{The small velocity limit}
\new{Let us study \eqref{eq:FinsFriedVac} a bit further to get more insights on where does the variety of vacuum solutions come from.}

\new{Employing a power series Ansatz 
\begin{align}\label{eq:power}
    f(s) = \sum_{i=0}^\infty f_i s^i\,,
\end{align}
we determine the coefficients $f_i$ order by order. To lowest orders in $s$, i.e.\ in the small speed approximation (stemming from the expansion of $\mathcal{G}_{ \mathcal{H}}$, see Appendix \ref{app:G})
\begin{align}
    \frac{3 f_1{}^2 H(\t)^2}{16 f_0 f_2{}^3 s^4} + \mathcal{O}(s^{-3}) = 0\,,
\end{align}
which immediately forces us to set $f_1 = 0$ in order to have a smooth $s\to 0$ limit. After setting $f_1=0$, the remaining term which contains  a negative power of $s$ (arising from the expansion of $\mathcal{G}_{ \dot{\mathcal{H}}}$) in \eqref{eq:FinsFriedVac} is
\begin{align}
    \frac{3 f_3 \dot{\mathcal{H}}(\t)}{2 f_0 f_2{}^2 s} + \mathcal{O}(s^0) = 0\,,
\end{align}
which implies $f_3=0$. Having identified these two necessary constraints, the next non-vanishing order is the $s^0$ order
\begin{align}\label{eq:s0}
    \frac{3 }{f_0{}^2} \dot{\mathcal{H}}(\t)+\left(\frac{30 f_4}{f_0 f_2{}^2}+\frac{9}{f_0{}^2}\right) \mathcal{H}(\t)^2+\frac{3 k}{f_0 f_2} + \mathcal{O}(s)= 0\,.
\end{align}
}
\new{The $s^1$ order for example yields
\begin{align}
    s f_5 \left(\frac{225   }{f_0 f_2{}^2} \mathcal{H}(\t)^2-\frac{45   }{2 f_0 f_2{}^2} \dot{\mathcal{H}}(\t) \right)= 0\,,
\end{align}
while the $s^2$ order gives,
\begin{align}\label{eq:s2}
    0=s^2\bigg(&
    \left(\frac{12 f_2 }{f_0{}^3}+\frac{25 f_4 }{f_0{}^2 f_2}-\frac{160 f_4{}^2}{f_0 f_2{}^3}+\frac{105 f_6}{f_0 f_2{}^2}\right) \dot{\mathcal{H}}(\t)\\
    &+\left(\frac{1960 f_4{}^2 }{f_0 f_2{}^3}+\frac{48 f_2 }{f_0{}^3}
    +\frac{155 f_4 }{f_0{}^2 f_2}-\frac{945 f_6 }{f_0 f_2{}^2}\right)\mathcal{H}(\t)^2\nonumber \\
    &+  \left(\frac{10 f_4 }{f_0 f_2{}^2}+\frac{11 }{f_0{}^2}\right)k
    \bigg)\,.
\end{align}
Following the general discussion in the beginning of Section \ref{ssec:vacsol}, these equations are of the form of \eqref{eq:dH=}, and the choice of the coefficients $c_1$ and $c_2$ implies relations between the expansion coefficients $f_i$.}

\new{The difference between these multiple Friedmann-type equations—which consistently emerge from the Finsler gravity vacuum equation \eqref{eq:FinsFriedVac}—and the vacuum Friedmann equations in general relativity,
\begin{align}\label{eq:GRFried}
\frac{\mathcal{H}^2}{a^2}  = - \frac{k}{a^2}\,,\quad
\frac{\dot{\mathcal{H}}}{a^2} = 0\,,
\end{align}
lies in the fact that the causal structure $f(s)$ is not fixed a priory as in general relativity, but must be determined from \eqref{eq:s-dep1} and \eqref{eq:s-dep2}. As a result, the terms proportional to $\dot{\mathcal{H}}$ and $\mathcal{H}^2$ do not necessarily decouple as in \eqref{eq:GRFried}. In vacuum general relativity, the only way to avoid this decoupling (which famously enforces the trivial, Minkowski spacetime solution) is to add a cosmological constant term which couples the two Friedmann vacuum equations as
\begin{align}
    \frac{\mathcal{H}^2}{a^2}  &= - \frac{k}{a^2} + \frac{\Lambda}{3} \,,\quad
    \frac{\dot{\mathcal{H}}}{a^2} = \frac{\Lambda}{3} \,,
\end{align}
resulting in the necessary condition
\begin{align}
    \mathcal{H}^2 + k - \dot{\mathcal{H}} = 0\,,
\end{align}
which is of the type \eqref{eq:dH=} with $c_1=1$ and $c_2=k$. In contrast, in the Finsler case, the choice of the coefficients $f_4$ in \eqref{eq:s0} and $f_6$ in \eqref{eq:s2} control to lowest order if there exists a solution to these equations, so that we obtain decoupled equations for $\dot{\mathcal{H}}$ and $\mathcal{H}^2$. If these coefficients are not finetuned, generically, the terms $\dot{\mathcal{H}}$ and $\mathcal{H}^2$ do not decouple. This feature can also be seen in the all order equation resulting in \eqref{eq:dH=}, a decoupled condition for  $\dot{\mathcal{H}}$ and $\mathcal{H}^2$ is generically not realized, but only for $c_1=0$. This then implies that $\mathcal{G}_{\mathcal{H}} = 0$, which is a constraint on the function $f(s)$.}

\new{Next we demonstrate explicitly for the exponentially expanding universe -- that is, $c_1=1, c_2=k=0$ -- that there exists a viable causal structure function $f(s)$ such that the full Finsler-Friedmann equation \eqref{eq:FinsFriedVac}, decaying into \eqref{eq:dH=}, \eqref{eq:s-dep1} and \eqref{eq:s-dep2} is solved, without a decoupling of the $\dot{\mathcal{H}}$ and $\mathcal{H}^2$ term.}

\subsubsection{The causal structure of the exponentially expanding vacuum universe}
Equations \eqref{eq:s-dep1} and \eqref{eq:s-dep2}, determine the function $f(s)$ and with it, the causal structure of spacetime. In contrast to the time evolution equations, which could be solved in conformal time in all generality, we are not able to do the same for these sixth order nonlinear ordinary differential equations for $f.$ 

In order to find a physically interesting and viable solution, that can be compared to classical FLRW cosmology, we focus on the case $c_2=k=0$. The first choice $c_2=0$ is motivated from our finding of the accelerated expanding vacuum universe \eqref{eq:expuni} and the second one, by the current observational constraint that the spatial curvature $k$ of the universe is very close to zero. 

For this choice of parameters, equation \eqref{eq:s-dep1} becomes an identity, thus it only remains to solve equation \eqref{eq:s-dep2}. As this is, still, a 6-th-order ordinary differential equation, we are not able to find a solution in closed form in terms of elementary functions. Therefore, we will proceed perturbatively and numerically to find the \emph{exponentially expanding Finsler FLRW}  spacetime geometry.

Taking a power series ansatz in $s$, as in \eqref{eq:power} we determine the first six non-vanishing coefficients of $f(s)$, which then serve as initial conditions to numerically determine the solution to all orders.

An immediate observation from using \eqref{eq:power} in \eqref{eq:s-dep2} is that the coefficients $f_1$ and $f_3$ must vanish. \new{In order to ensure time reversal symmetry we assume that $f$ does not contain any odd powers of $s$.} Hence, a refined ansatz is
\begin{align}
    f(s(X)) =  \sum_{i=0}^\infty f_{2i} S^{2i} = \sum_{i=0}^\infty Q_{i} X^{i} = Q(X(s)) \,.
\end{align}
Then, it turns out that the coefficients $Q_0$ and $Q_1$ can be chosen freely, while the next four coefficients $Q_2$ to $Q_5$ are obtained by solving equation \eqref{eq:s-dep2} order by order in $X$,
\begin{align}\label{eq:bdry}
    Q_2 &= -\frac{(c_1+3) Q_1^2}{10 Q_0}\\
    Q_3 &= \frac{\left(16 c_1^3-75 c_1^2-922 c_1-1779\right) Q_1^3}{1050 (c_1-9) Q_0^2}\\
    Q_4 &= -\frac{\left(490 c_1^5-6359 c_1^4-22227 c_1^3+321773 c_1^2+1778345 c_1+2576298\right) Q_1^4}{189000 \left(c_1^2-19
   c_1+90\right) Q_0^3}\\
    Q_5 &=\frac{Q_1^5}{363825000(c_1-9)^2 \left(3 c_1^2-65 c_1+350\right) Q_0^4}\left( \right.511320 c_1^8-16231154 c_1^7 +125704897 c_1^6+ \nonumber\\
    &  \left.+582150849 c_1^5-23003963752 c_1^3 +214489583169 c_1^2 +1071379945593 c_1 +1431584382546\right)\,.
\end{align}
To determine suitable values for the coefficients $Q_0$ and $Q_1$ we compare the power series expansion of $f(s)$ to their values in FLRW geometry, for which
\begin{align}\label{eq:FLRW}
    f(s) = \sqrt{1-s^2} \sim 1 - \frac{1}{2}s^2 + \mathcal{O}(s^4)\,.
\end{align}
Therefore, in order to obtain a causal structure which for slow velocities looks like FLRW geometry, we choose
\begin{align}\label{eq:Q0-Q1}
    Q_0 = 1\,,\quad Q_1 = -\frac{1}{2}\,.
\end{align}
We like to point out that deviations in the causal structure generically appear from order $s^4$ on; a peculiar case is $c_1=-3$, which would lead to a causal structure which differs from FLRW geometry only at order $s^6$. However, the most interesting case, from our point of view, is $c_1=1$, since it directly leads to an exponentially expanding universe as we saw in \eqref{eq:expuni}.

\begin{figure}[h!]
    \centering
    \includegraphics[width=0.7\linewidth]{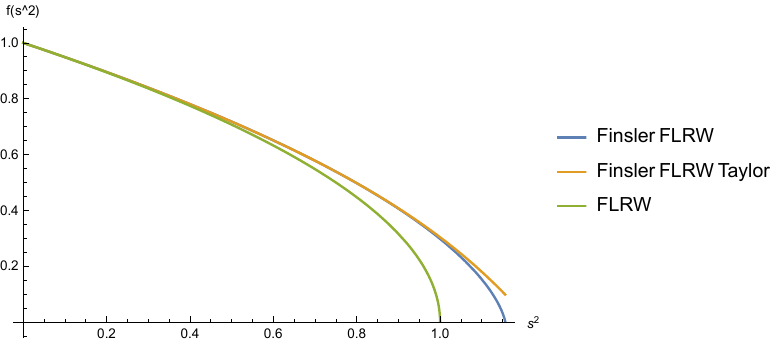}
    \caption{Numerical solution of equation \eqref{eq:s-dep2}, with boundary conditions \eqref{eq:bdry} and $c_1=1$ (blue line) compared to the approximate Taylor series solution up to $s^{20}$ with $c_1=1$ (orange line) and classical FLRW geometry (green line.)}
    \label{fig:numsol}
\end{figure}

\begin{figure}[h!]
    \centering
    \includegraphics[width=0.4\linewidth]{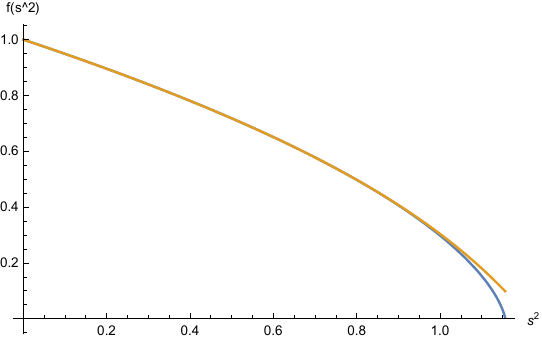}
    \includegraphics[width=0.59\linewidth]{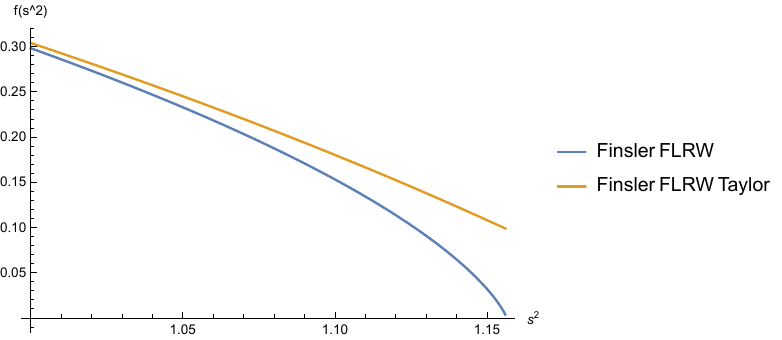}
    \caption{Numerical solution of equation \eqref{eq:s-dep2}, with boundary conditions \eqref{eq:bdry} and $c_1=1$ (blue line) compared to perturbative Taylor expansion solution of equation \eqref{eq:s-dep2} (orange line). Left: Whole range. Right: Zoomed in.}
    \label{fig:numsol3}
\end{figure}

The first six coefficients in the power series expansion $Q_0$ to $Q_5$ are used as boundary conditions to solve equation \eqref{eq:s-dep2} numerically. For $c_1=1, Q_0 = 1$ and $Q_1 = -\frac{1}{2}$ we have that
\begin{align}\label{eq:Q2-Q5}
    Q_2 = -\frac{1}{10},\quad 
    Q_3 = -\frac{23}{560},\quad 
    Q_4 = -\frac{269}{12600},\quad 
    Q_5 = -\frac{97167659}{7761600000}\,.
\end{align}
Employing the $NDSolve$ algorithm of Mathematica \cite{mathematica}, with these initial conditions, leads to the \emph{Finsler FLRW} solution displayed in Figure \ref{fig:numsol}, where we compare the all order numerical solution for $f(s^2)$ with the approximate Taylor series solution and up to order $s^{20}$ and the non-approximate FLRW geometry given by \eqref{eq:FLRW}.
\noindent The difference between the numerical and the perturbative solution up to order $s^{20}$ is shown in Figure \ref{fig:numsol3}.

\begin{figure}[h!]
    \centering
    \includegraphics[width=0.4\linewidth]{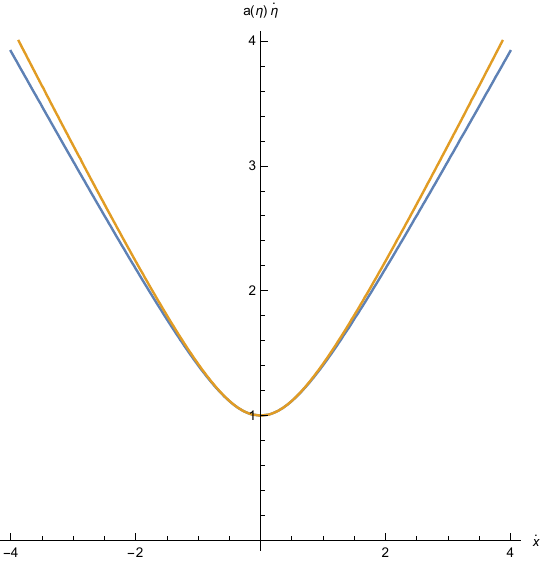}\qquad \quad
    \includegraphics[width=0.51\linewidth]{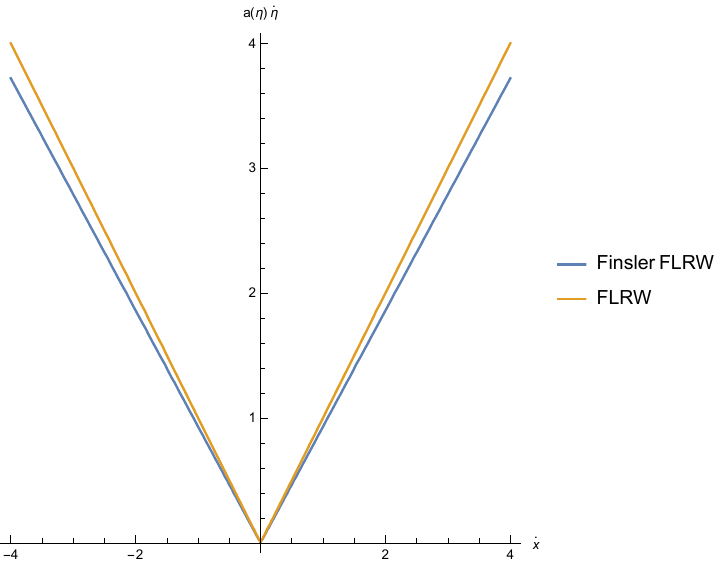}
    \caption{Left: The future pointing unit timelike directions $L=1$ of the Finsler FLRW geometry (blue) compared to FLRW geometry (orange), with two spatial directions suppressed. Right: The  corresponding lightcones. For small values of $\dot x$ the causal structures are nearly undistinguishable. Differences only appear for larger $\dot x$, respectively for larger values of the symmetry adapted variable~$s$.}
    \label{fig:Causal}
\end{figure}

The point $s_0$ where $f(s_0)=0$ marks the lightcones of the geometry, as is clear from the conformal ansatz of our geometry \eqref{eq:LHomIso}. While for FLRW geometry in this parametrization, this is given by $s_0=\pm 1$, for the numerical solution the lightcone is reached by $s_0\sim \pm 1.07517$. Thus, the lightcones in the Finslerian geometry are wider as displayed on the right in Figure \ref{fig:Causal}. \new{This feature has direct consequences for the famous distance to redshift relation. Since light rays propagate on curves satisfying $s= s_0$ and $s = w/\dot\t$, for radial light rays ($w=\dot r/\sqrt{1-kr^2}$, see \eqref{eq:homisoL}) the relation
\begin{align}
    s = \frac{w}{\dot \t} = \frac{a w}{\dot t} = \frac{\dot r}{\sqrt{1-kr^2}} \frac{a}{\dot t} = \frac{dr}{dt} \frac{a}{\sqrt{1-kr^2}} = s_0\,,
\end{align}
holds in cosmological time. For FLRW geometry the constant $s_0$ carries the value $1$, while for our Finsler modification this value is slightly larger. The conversion of this expression from cosmological time $t$ to redshift $z$ is modified on Finsler spacetimes, compared to pseudo-Riemannian General Relativity \cite{Hasse:2019zqi} and must be investigated carefully. Some further phenomenological consequences following from modified distance to redshift relation due to Finsler spacetime geometry have been discussed in \cite{Hohmann:2016pyt}. A detailed study of the impact of the modified causal structure will be performed in future works.}

In the range we are interested in, $s\in (0,1.07517)$, the numerical all-order solution for $f(s)$ and the approximate Taylor series solution are in close agreement with each other, and nearly indistinguishable for $s<1$. Hence, we use the approximate solution to display the future-pointing unit timelike cones on the left of Figure \ref{fig:Causal}.
Similarly to the lightcones, we find that the set of all unit timelike directions is opened more widely, compared to FLRW geometry. It is no longer a hyperboloid, but a deformation thereof. Near $s=0$, it is close to the hyperbolic shape of FLRW geometry and, the larger the value of $s$, the more different these shapes become. Physically speaking, this means observers (massive objects) propagating with a trajectory whose tangent is close to the canonical cosmological time direction $V = \partial_t$ see basically a FLRW geometry, while only observers propagating with a spatial velocity large against $V$ experience relativistic effects that deviate from pseudo-Riemannian geometry.

At this point, we have found a Finsler modification of FLRW geometry which represents the gravitational field of the vacuum of a kinetic gas universe. Its causal structure is undistinguishable from FLRW geometry for small  spatial velocities $s$ of massive objects / observers, relative to the canonical cosmological propagation direction $V=\partial_t$ - and  only differs from FLRW geometry for massive objects propagating on trajectories that correspond to a large $s$. While the difference between the causal structures is rather mild, the difference in the evolution of the universe is striking.

The gravitational degrees of freedom, which only become visible from the phase space (tangent bundle/cotangent bundle) point of view, that are sourced by the 1PDF of a kinetic gas, lead naturally to an exponential expansion of the universe, which quantifies the conjecture we made in \cite{Hohmann:2020yia} for the first time: the usually neglected parts of the 1PDF of a kinetic gas are a source of dark energy.

\section{Discussion}\label{sec:disc}
The origin of dark matter and dark energy belongs to the most puzzling questions in modern fundamental physics. Even without knowing their origin, the so far most successful theoretical model $\Lambda CDM$, which assumes a cosmological constant and a specific cold form of dark matter, is currently challenged by the discrepancies in the measurements of the value of the "present-day" Hubble function and other cosmological parameters, \cite{Abdalla:2022yfr,DiValentino:2025sru}.

In the $\Lambda CDM$ model, the matter content of the universe as a whole is described in terms of a (perfect) fluid energy-momentum tensor, \new{supplemented by an equation of state}. It determines the cosmological evolution of the universe through the Einstein equations. The (perfect) fluid energy-momentum tensor is, however, only an effective description of a multi-particle system, not capturing all of its properties. \new{Thus, when the energy–momentum tensor is used as the source in the Einstein equations, it yields only a partial description of the system’s gravitational field—specifically, the parts generated by the information about the system contained in the tensor.}

In this article we discussed and demonstrated explicitly how the gravitational field of a physical multi-particle system can be derived from a more fundamental point of view, namely, instead of from its energy-momentum tensor through the Einstein equations, directly from its 1PDF, which acts as a source of the Finsler gravity equation. Since the 1PDF of a kinetic gas contains significantly more information about the system than the corresponding energy-momentum tensor (which is typically derived through an averaging procedure), we quantified how these additional features of the gas -- particularly its velocity distribution -- contribute to its gravitational field. The consequences of this new point of view are striking. 

When applied to homogeneous and isotropic symmetry, and considering such Finsler spacetime geometries for which the existence of a conformal cosmological time is ensured, we found the \emph{Finsler-Friedmann equation} \eqref{eq:FinsFried}, that has the 1PDF of a homogeneous and isotropic kinetic gas as a source term. We discussed in detail how the Finsler-Friedmann equation on phase space takes \emph{all} the properties of the gas that are encoded in the 1PDF into account, when determining the gravitational field. Moreover, we showed that this phase space equation is in general equivalent to infinitely many tensor density equations on spacetime \eqref{eq:FriedTensDen} and not only to a 2-tensor equation as the Einstein equations are. This comparison makes it very visible that one neglects important aspects of the kinetic gas when deriving its gravitational field through the Einstein equations from an energy-momentum tensor. Namely, only specific parts of the velocity distribution of matter are taken into account, although its full velocity distribution is available. More specifically, in the Einstein equations only the second moment of the 1PDF is \new{directly} taken into account to derive the gravitational field of the gas. In the Finsler gravity equation, or the corresponding infinitely many effective gravitational field equations on spacetime, all moments of the 1PDF are taken into account. No moments are lost -- and all  of them contribute to the gravitational field.

The deeper physical meaning behind the phase space / tangent bundle dynamics of the 1PDF of the kinetic gas coupled through the Finsler gravity equation to the Finsler geometry of spacetime is that it not only determines the evolution of the universe in time, but also its causal structure, meaning the precise shape of the convex cones of past and future pointing unit-timelike directions. Mathematically, they are encoded in the direction/velocity dependence of the geometry defining a Finsler Lagrangian $L$. These degrees of freedom of the geometry of spacetime are fixed in the Einstein equations by assuming pseudo-Riemannian spacetime geometry; in contrast, in the Finsler gravity setting, they are subject to dynamical equations  \cite{Pfeifer:2011tk,Pfeifer:2011xi,math-foundations,Hohmann:2019sni}. In this way, we presented how to realize Wheeler's famous quote 
\begin{quote}
    "Space-time tells matter how to move; matter tells space-time how to curve"\,,
\end{quote}
for kinetic gases and their 1PDFs.

In homogeneous and isotropic symmetry, which we investigated, the causal structure of spacetime is determined from the $s = w/\dot t$ dependence. We found in Equation \eqref{eq:splitting-eta-s} that the Finsler-Friedmann equation allows us to partially separate the time evolution (the dependence of the spacetime geometry on conformal time $\t$), and the causal structure (the dependence of the spacetime geometry on $s$). Even more, in vacuum, i.e., for a vanishing 1PDF, the time evolution of the scale factor and the causal structure determination decouple completely. The former is determined by Equation \eqref{eq:dH=}, while the latter is determined by the Equations \eqref{eq:s-dep1} and \eqref{eq:s-dep2}.

The time evolution equation can be solved analytically in conformal time in all generality. The most exciting finding is that there exists a choice of integration constants such that the corresponding cosmological time evolution gives an exponential expansion of the universe \eqref{eq:expuni}, without the need of a cosmological constant, that is, without the need of dark energy. This is a purely geometric finding for the vacuum state of the kinetic gas spacetime geometry. We thus found the first direct evidence for the conjecture that the dark energy phenomenon \new{can}, at least partly, \new{be explained} by \new{a Finslerian spacetime geometry that is capable to include} the contribution of the velocity distribution of many particle systems to their gravitational field.

For the moment, we were not able to solve the corresponding dynamical equation for the causal structure \eqref{eq:s-dep2} analytically. Yet, we could find a spacetime solution numerically and perturbatively. Both the numerical and the perturbative solutions (see Figure \ref{fig:numsol}) show a behavior that deviates from pseudo-Riemannian FLRW geometry, but in a very controlled way. The deviation is negligible around the canonical cosmological time direction, i.e.\ for massive objects at rest or with small velocities. The deviation increases the faster the objects are, the closer they propagate with a velocity near the speed of light, or near the lightlike directions. This finding allows the interpretation that \emph{pseudo-Riemannian geometry sourced by an energy-momentum tensor is a small velocity approximation of a more general Finslerian spacetime structure sourced by a kinetic gas.} 

\section{Conclusion and Outlook}\label{sec:conc}

In conclusion, we found that taking the full 1PDF of kinetic gases as source of their gravitational field, requires an extended notion of spacetime geometry beyond pseudo-Riemannian geometry. The most natural candidate to derive the contribution of all moments of the 1PDF to the gravitational field of the gas is Finslerian spacetime geometry. Already in vacuum, Finsler geometry gives rise to at least one exponentially expanding universe as a solution. This finding does not require any exotic form of matter or energy, it just requires that we use the spacetime geometry that can take into account as many properties of the matter as possible. Here, this means not fixing the shapes of the sets of causal directions (lightcones and unit-timelike directions) \emph{a priori}, but rather allowing them to be dynamically determined by the source matter. \new{We already outlined consequences of this modifications in the causal structure for the distance to redshift relation. A systematic study of phenomenological observable consequences is an important future investigation of Finsler spacetime geometry as geometry of spacetime sourced by kinetic gases.}

The very promising vacuum solutions, whose time volution we completely classified, are only the first step towards a realistic kinetic gas Finsler spacetime description of our universe. \new{On the one hand, we need to study if all of the resulting scale factors lead to physically viable Finsler spacetimes, or which of them need to be discarded as unphysical.} \new{On the other hand, the next} important step is to find a realistic ansatz for the 1PDF that describes the cosmological kinetic gas, and to solve the Finsler-Friedmann equation for this case, at least, numerically. The question is, how does a realistic 1PDF of the kinetic gas that fills the universe look like?

We expect that, just as for the vacuum case, solutions to the non-vacuum Finsler-Friedmann equation will naturally lead to an exponential expansion of the universe, that can reduce the need for dark energy, or even make it superfluous. In both cases, the answer to what produces dark energy would be the forgotten moments of kinetic gas matter.


\acknowledgments
The authors would like to acknowledge networking support by the COST Actions CA23130 “Bridging high and low energies in search of quantum gravity (BridgeQG)” and CA21136 "Addressing observational tensions in cosmology with systematics and fundamental physics (CosmoVerse)".  CP acknowledges support by the excellence cluster QuantumFrontiers of the German Research Foundation (Deutsche Forschungsgemeinschaft, DFG) under Germany's Excellence Strategy -- EXC-2123 QuantumFrontiers -- 390837967 and was funded by the Deutsche Forschungsgemeinschaft (DFG, German Research Foundation) - Project Number 420243324 - and by the Transilvania Fellowships for Visiting Professors grant 2024 of the Transilvania University of Brasov.

\appendix

\section{Some basic notions of Finsler geometry}\label{app:FinsGeom}
In this Appendix, we briefly introduce the Finsler geometric objects that represent the building blocks of the Finsler gravity equation \eqref{eq:homisoL}. For further details, we refer to the textbooks \cite{BCS,Bucataru} or the articles \cite{math-foundations,Pfeifer:2019wus}.
\begin{itemize}
    \item The Cartan tensor and its trace:
	\begin{align}\label{eq:Cartan}
		C_{abc} = \frac{1}{4}\dot{\partial}_a \dot{\partial}_b \dot{\partial}_c L = \frac{1}{2} \dot{\partial}_c g_{ab}, \quad  C_a = g^{bc}C_{abc}\,.
	\end{align}
        The Cartan tensor measures, at each point $x \in M$, how much the Finslerian metric tensor $g_{ab}(x,\dot{x})$ differs from a pseudo-Riemannian (i.e., quadratic in $\dot{x}$) one.
    \item The geodesic spray coefficients $G^a$, the canonical nonlinear connection coefficients $N^a{}_b$ and the horizontal derivatives $\delta_a$:
	\begin{align}\label{eq:geodspray}
		G^a = \frac{1}{4} g^{ab}(\dot x^c \partial_c \dot \partial_b L - \partial_b L)\,,\quad N^a{}_b = \dot{\partial}_b G^a\,,\quad \delta_a = \partial_a - N^b{}_a \dot\partial_b\,.
	\end{align}
        \item The curvature of the canonical nonlinear connection and the canonical curvature scalar (or \textit{Finsler-Ricci} scalar:
	\begin{align}\label{eq:curv}
		R^c{}_{ab}\dot \partial_c = [\delta_a,\delta_b]\,,\quad R = R^a{}_{ab}\dot x^b \,.
	\end{align}
        \item The Chern-Rund covariant derivatives, defined on the \emph{adapted basis}  $\{\delta_a,\dot{\partial}_a\}$ of $T_{(x,\dot x)}TM$ as:
	\begin{align}
		\nabla_{\delta_{a}}\delta_{b}&=\Gamma^{c}{}_{ab}\delta_{c}=\frac{1}{2}g^{cd}\left(\delta_{a}g_{bd}+\delta_{b}g_{ad}-\delta_{d}g_{ab}\right)\delta_{c}\,, &  & \nabla_{\delta_{a}}\dot{\partial}_{b}=\Gamma^{c}{}_{ab}\dot{\partial}_{c}\,,\\
		\nabla_{\dot{\partial}_{a}}\delta_{b}&=0, &  & \nabla_{\dot{\partial}_{a}}\dot{\partial}_{b}=0\,.
	\end{align}
        We note that, in the particular case of pseudo-Riemannian geometry, $\Gamma^{c}{}_{ab}$ become the usual Christoffel symbols, whereas the Finsler-Ricci scalar $R=R^{}{^a}_{bac}\dot{x}^b\dot{x}^c$ is obtained from the usual Ricci tensor by contracting it twice by $\dot{x}$.
        \item The dynamical covariant derivative - measuring the rate of variation of tensor fields along geodesics of $(M,L)$,  is obtained from the Chern-Rund one by contraction with $\dot x^a$ and expressed by the symbol $\nabla:=\dot x^a \nabla_{\delta_{a}}$.
	\item The Landsberg tensor and its trace
	\begin{align}\label{eq:lands}
		P_{abc} = \nabla C_{abc}\,,\quad P_a = g^{bc}\nabla C_{abc}\,,
	\end{align}
which keep track of how much the "non-Riemannianity" of our Finsler space varies as one moves along its geodesics.
\end{itemize}

\section{Homogeneous and isotropic Finsler geometry}\label{app:HomAndIsoGeom}
In this appendix, we explicitly list the geometric tensors of Finsler geometry in homogeneous and isotropic symmetry. They were obtained in \cite{Friedl-Szasz:2024vtu} and serve as the foundation for deriving the conformal homogeneous and isotropic Finsler-Friedmann equation \eqref{eq:FinsFried}. We first display them here for general homogeneous and isotropic symmetry, for further use in the future. Then, for the derivations in this article, we set $t\to \t$ and $h(t,s) \to a(\t) f(s)$.

\subsection{Finsler metric and Cartan tensor}\label{sapp:HomIsoCartan}
We begin with the Finsler metric and the Cartan tensor for cosmologically symmetric Finsler Lagrangians $L$, which are all described by:
\begin{align}
	L = \dot t^2 h(t,s)^2\,, \quad s=\frac{w}{\dot t}\,,\quad w^2 = \frac{\dot r^2}{1-k r^2}+r^2\big(\dot \theta^2+\dot\phi^2\sin^2\theta \big)\,.
\end{align}
We note that $w^2$ is the Finsler function of the metric on the spatial slices $t=const.$, which, just as in the FLRW case, is actually a 3-dimensional Riemannian metric of constant scalar curvature (the difference, in the Finslerian case, is the much more general intertwining between space and time, encoded in the general, non-quadratic dependence of $h$ on $s$). Introducing the notations,
\begin{align}
w_{\alpha\beta} = \frac{1}{2}\dot\partial_\alpha \dot\partial_\beta w^2
= \textrm{diag}\left(\frac{1}{1-kr^2},r^2,r^2 sin^2\theta\right)\,,
\quad
\dot{\partial}_{\alpha} w = w_\alpha = \frac{1}{w} w_{\alpha\beta}\dot x^\beta\,,
\quad
w_a = (-s, w_\alpha)\,,
\end{align}
we first obtain:
\begin{align}
	\dot{\partial}_0 s = -\frac{s}{\dot t}\,,\quad
	\dot{\partial}_\alpha s = \frac{w_\alpha}{ \dot t}\,.
\end{align}
Denoting by primes derivatives with respect to $s$, we find:
\begin{itemize}
    \item The components of the Finsler metric and its inverse:
    \begin{align}\label{metric_comps_cosmo}
	g_{00}&=h^{2}-2shh'+s^{2}(h'{}^{2}+h h'')\,,\\
	g_{0\alpha}&=w_{\alpha}\left(h h'-s(h'{}^{2}+h h'')\right)\,,\\
	g_{\alpha\beta}&=h h' \frac{w_{\alpha\beta}}{s}+\left(s h'{}^{2}+s h h''-h h'\right)\frac{w_{\alpha}w_{\beta}}{s}\,,
    \end{align}
    and
    \begin{align} \label{eq:inverse_metric}
	g^{00}&=A\,, &  A&=\frac{h'^2+ h h''}{h^3 h''}\,, \nonumber\\
	g^{0\alpha }&=B\frac{\dot{x}^{\alpha }}{w}\,, & B&=\frac{s (h'^2+ h h'') - h h'}{h^3 h''}\,,\\
	g^{\alpha \beta }&=C w^{\alpha \beta }+\frac{D}{w^{2}}\dot{x}^{\alpha }\dot{x}^{\beta }\,, & C&=\frac{s}{ hh{' }}\,, D=\frac{(h-sh')\left(hh'-s(h'{}^{2}+hh'')\right)}{h^{3}h'h''}\,,\nonumber
    \end{align}
    where $w^{\alpha\beta}=\text{diag}(1-kr^2,\frac{1}{r^2},\frac{1}{r^2sin^2\theta})$  is the inverse of $w_{\alpha\beta}$.
    \item The Cartan tensor:
    \begin{align}
	C_{000} &= -\frac{1}{2} \frac{s^3}{\dot t} ( h h''' + 3 h' h'')\,,\\
	C_{00\alpha} &= w_\alpha \frac{s^3}{2 w} ( h h''' + 3 h' h'')\,,\\
	C_{0\alpha\beta}
        &= \frac{w_{\alpha\beta}}{2 w}\left( h (h'-s h'')-  s h'^2 \right) - \frac{w_\alpha w_\beta}{2 w}\left( h(h'-sh'')- s h'^2 + s^2 (h h''' + 3 h' h'') \right)\,,\\
	C_{\alpha\beta\gamma} &= - \frac{(w_{\beta\gamma} w_{\alpha} + w_{\alpha\gamma} w_{\beta} + w_{\alpha\beta}        w_{\gamma} - 3 w_{\alpha} w_{\beta} w_{\gamma}) \left( h (h' -  s h'') - s h'^2\right)}{2 s w} \nonumber\\
	&+ \frac{s w_{\alpha} w_{\beta} w_{\gamma} ( h h''' + 3 h' h'')}{2 w} \,.
    \end{align}
    Introducing the scalar variable $T = \dot t w$, this can be combined to a full Finsler spacetime tensor as
    \begin{align}
        C_{abc}
        &= \frac{1}{2} \left( h (h' -  s h'') - s h'^2\right) T_{abc} + \frac{\left( h h''' + 3 h' h''\right)s}{2 w} w_a w_b w_c\,,
    \end{align}
    where:
    \begin{align}
        T_{abc} &:= \dot\partial_a \dot\partial_b \dot\partial_c T\\
        &= (\delta_c^0 \delta_b^\beta \delta_a^\alpha  + \delta_a^0 \delta_b^\beta \delta_c^\alpha +\delta_b^0 \delta_c^\beta \delta_a^\alpha) \left(\tfrac{w_{\alpha\beta}-w_\alpha w_\beta}{w}\right)\nonumber\\
        &+ \frac{\dot t}{w^2}\delta_c^\gamma \delta_b^\beta \delta_a^\alpha (3 w_\alpha w_\beta w_\gamma - w_{\alpha\beta} w_\gamma - w_{\alpha\gamma} w_\beta - w_{\gamma\beta} w_\alpha)\,.
    \end{align}
\end{itemize}

\subsection{Geodesic spray components, nonlinear connection coefficients and Landsberg tensor}\label{sapp:HomIsoGandN}
We continue by deriving the geodesic spray and the canonical Cartan non-linear connection. Dots over functions will mean derivatives with respect to $t$, in order to distinguish them from derivatives w.r.t. $s$, which are denoted by primes. The dot over a function should not be confused with the symbols for the velocity coordinates $\dot x$ or the corresponding derivatives $\dot \partial$.

From the definition of the geodesic spray \eqref{eq:geodspray} we find:
\begin{align}
    G^0 = \dot t^2 \frac{\bigl(h'' \dot h -  h' \dot h'\bigr)}{2 h h''}
\end{align}
and
\begin{align}
    G^\alpha =   \tilde G^\alpha
    + \frac{1}{2} \dot t \dot{x}^{\alpha}
    \frac{ \Bigl(s h'' \dot h + \bigl(h -  s h'\bigr) \dot h'\Bigr) }{s h  h''}\,,
\end{align}
where the quantities $\tilde G^\alpha$ are the geodesic spray coefficients of the 3-dimensional Finsler function $w^2$, given by
\begin{align}
	\tilde G^\alpha = \frac{1}{4}w^{\alpha\beta}\left( \dot x^\gamma \dot{\partial}_\beta \partial_{\gamma} w^2- \partial_\beta w^2\right)\,.
\end{align}
The nonlinear connection coefficients are then obtained as $N^a{}_b = \dot \partial_b G^a$, as follows:
\begin{align}
    N^0{}_0
    &= \frac{\dot t}{2 h^2 h''^2}\left[ h''^2(sh'+2h)\dot h - h'\bigl((sh'+2)h''+shh'''\bigr) \dot h'+sh'h''\dot h''\right]\,,\\
    N^0{}_\alpha
    &= \frac{\dot t w_{\alpha} h'}{2 s h^2 h''^2}  \left[ -h''^2\dot h + (h'h''+hh''')\dot h'-hh''\dot h''\right]\,,\\
    N^\alpha{}_0
    &= \frac{\dot{x}^{\alpha}}{2 h^2 h''^2}
    \biggl[ sh''^2(h+sh')\dot h + \Bigl( h''\bigl( 2h^2-sh'(h+sh')\bigr)+shh'''(h-sh')\Bigr)\dot h'\nonumber\\
    &- sh^2h''(h-sh')\dot h''\biggr],\\
    N^\alpha{}_\beta
    &= \Gamma^{\alpha}{}_{\beta\gamma} \dot{x}^{\gamma}
    + \frac{\dot t \delta^{\alpha}_{\beta}}{2 s h h''} \left[s h'' \dot h + \bigl(h -  s h'\bigr) \dot h'\right] - \frac{w_{\beta} \dot{x}^{\alpha}}{2 s^2  h''^2} \left[(h''+sh''')\dot h'-s\dot h'' \right]  \nonumber\\
    &+ \frac{w_{\beta} \dot{x}^{\alpha} h'}{2 h^2  h''^2} \left[-h''^2\dot h + (h'h''+hh''')\dot h'-hh''\dot h''\right].
\end{align}
where $\Gamma^{\alpha}{}_{\beta\gamma}$ are the coefficients of the Levi-Civita connection of the spatial metric with components $w_{\alpha\beta}$.

We note that the dynamical covariant derivative acts on functions $\varphi(t,s)$ of $t$ and $s$ as
\begin{align}\label{eq:nablah}
    \nabla \varphi(t,s) = \dot t\left( \partial_t \varphi - \frac{\dot h'}{h''} \partial_s \varphi\right)\,.
\end{align}

The Landsberg tensor \eqref{eq:lands} can now be displayed as
\begin{align}\label{eq:landshomiso}
    4 P_{abc} 
    &=  \left( \nabla\left( h (h' -  s h'') - s h'^2\right) + \dot t q(t,s) \left( h (h' -  s h'') - s h'^2\right)  \right)T_{abc} \\
    &+ \left(\nabla \left(  \tfrac{( h h''' + 3 h' h'')}{\dot t}\right) + 3 \left( h h''' + 3 h' h''\right)p(t,s)   \right)w_a w_b w_c\,,
\end{align}
where we used that 
\begin{align}\label{eq:nablaw}
    \nabla w_a &= p(t,s) w_a \dot t\,,\quad \nabla T_{abc} = q(t,s) \dot t T_{abc}\,,
\end{align}
with
\begin{align}
    p(t,s) = \frac{1}{2 h h''^2}\bigl[ -h''^2\dot h + (h'h''+hh''')\dot h'-hh''\dot h''\big]\,,
\end{align}
and
\begin{align}
    q(t,s) =  \frac{1}{2 sh h''^2}\Bigl[ -sh''^2\dot h+\bigl( h''(2h+sh')+shh'''\bigr)\dot h'-shh''\dot h''\Bigr]\,.
\end{align}

\subsection{The curvature}\label{app:HomAndIsoR}
The last ingredient to evaluate the Finsler-Friedmann equation \eqref{eq:FinsFried} is the Finsler-Ricci curvature scalar $R$, given by
\begin{align}
    & \frac{R}{w^2} 
    =   \frac{R^a{}_{ab}\dot x^b }{w^2}
    = -2 k 
    -  \frac{3  h'''^2 \dot h'^2}{s^2 4 h''^4} 
    -  \frac{9  \bigl(h'' \dot h -  h' \dot h'\bigr)^2}{4h^2 s^2  h''^2} \\
    &+ \frac{  \dot h' \Bigl(s h'''' \dot h' + h''' \bigl(2 \dot h' + 3 s \dot h''\bigr)\Bigr)}{2s^3 h''^3} 
    + \frac{   2 \ddot h' + s \ddot h''}{2s^3  h''}\\
    &+ \frac{3   \biggl(- h' h''' \dot h'^2 + 2 h' h'' \dot h' \dot h'' + h''^3 \ddot h -  h''^2 \Bigl(\dot h'^2 + h' \ddot h'\Bigr)\biggr)}{2h s^2  h''^3}\\ 
    &-  \frac{ -6 \dot h'^2 + 4 s \dot h' \bigl(2 \dot h'' + s \dot h'''\bigr) + s^2 \Bigl(\dot h''^2 + 2 h''' \ddot h'\Bigr)}{4s^4  h''^2}\,.
\end{align}

\section{The coefficients of the Finsler gravity equations}\label{app:G}
We display the explicit form of the coefficient functions introduced in \eqref{eq:FinsFried}. They are obtained from the geometric objects introduced in the previous Appendix \ref{app:HomAndIsoGeom} with help of the computer algebra program xAct \cite{xAct} and by redenoting $t$ by $\t$ as well as $h(t,s)$ by $a(\t)f(s)$. The notation is that $'$ denotes the derivative of $f$ with respect to $s$.

The coefficient function in front of the spatial curvature $k$ is given by
\begin{align}
    \mathcal{G}_k = \frac{2 \bigl(f -  s f'\bigr) \Bigl(- s f'^2 + f \bigl(f' + 2 s f''\bigr)\Bigr)}{f^3 f' f''}\,.
\end{align}

The coefficient function in front of the square of the conformal time Hubble function $\mathcal{H^2}$ is
\begin{align}
    \mathcal{G}_{\mathcal{H}} 
    &=\frac{3 }{4}\frac{f'^4(4ff'f'''s-15ff''^2s+2sf'^2f''+4ff'f'')}{sf^5f''^4}\\
    &+\frac{1 }{4}\frac{f'^2(-98s^2f''^4+4sf'f''^2(11f''+13sf''')-6f''^2-18s^2f'''^2-sf''(16f'''-7sf''''))}{s^2f^3f''^5}\\
    &+\frac{1}{2}\frac{33sf''^2-26sf'f''-22f'f''}{sf^2f''^2}+\frac{1}{4}\frac{f'^2(14f''^2+55s^2f'''^2+sf''(32f'''-27sf''''))}{s^2f^2f''^4}\\
    &-\frac{1}{4}\frac{f'^3(24s^2f'''^2-3sf''f'''(-4f'''+7sf'''')+f''^2(4f'''-4sf''''+3s^2f'''''))}{s^2f^2f''^6}\\
    &-\frac{1}{2}\frac{2s^2f''^4-2s^2f'f''^2f'''-11s^2f'^2f''f''''+17s^2f'^2f'''^2-2sf'f''^3-5f'^2f''^2}{s^2ff'^2f''^3}\\
    &+\frac{1}{2}\frac{f'(-8f''^3+27s^3f'''^3-28s^3f''f'''f''''+f''^2(-6sf'''+5s^3f'''''))}{s^3ff''^5}\\
    &+\frac{1}{4}\frac{f'^2(6f''^4-30s^4f'''^4+42sf''f'''^2f''''+f''^2(6s^2f'''^2-6s^4f''''^2-9s^4f'''f'''''))}{s^4ff''^7}\\
    &+\frac{1}{4}\frac{f'^2(8f'''-2sf''''+s^3f'''''')}{s^3ff''^4}.
\end{align}

And the coefficient function in front of  the conformal time derivative of the conformal time Hubble function $\dot{\mathcal{H}}$ is
\begin{align}
    \mathcal{G}_{\dot {\mathcal{H}}} 
    &=\frac{1}{4}\frac{f'^2(3sf'^2f''+sff'f'''-4sff''^2-8ff'f'')}{sf^4f''^3}\\
    &+\frac{1}{4}\frac{20sf'f''^3-6s^2f''^4-f'^2(2f''^2+s^2f'''^2+sf''(4f'''-sf''''))}{s^2f^2f''^4}\\
    &+\frac{1}{2}\frac{2s^2f''^4+s^2f'f''^2f'''-2s^2f'^2f''f''''+2s^2f'^2f'''^2-2sf'f''^2-2f'^2f''^2}{s^2ff'^2f''^3}\\
    &+\frac{1}{4}\frac{f'(4f''^3-3s^3f'''^3+4s^3f''f'''f''''+f''^2(2sf'''-s^3f'''''))}{s^3ff''^5}.
\end{align}

\section{The moment integrals}\label{app:IntInd}
The Finsler-Friedmann tensor-density equations \eqref{eq:FinsFriedInt} are constructed from integrals of the type 
\begin{align}\label{eq:int}
    \int_{\mathcal{S}_x} \left(\frac{\dot x^{a_1}...\dot x^{a_n}}{L^{n/2}} \varphi(\t,s) \dd\Sigma_x\right)\Big|_{\mathcal{S}_x}\,,
\end{align}
where the volume element is given by 
\begin{align}
    d\Sigma_x 
    &= \frac{-\det g}{L^2}\  \ i_{\mathbb{C}}\left(\dd \dot \t \wedge \dd \dot r \wedge \dd \dot \theta \wedge \dd \dot \phi\right)\\
    &= \frac{-\det g}{ a(\t)^4 f^4 }\ \ \frac{i_{\mathbb{C}}\left(\dd \dot \t \wedge \dd \dot r \wedge \dd \dot \theta \wedge \dd \dot \phi\right)}{\dot \t^4}\,,
\end{align}
and $\mathbb{C} = \dot x^a \dot \partial_a$ is the generator of the vector rescalings $\dot{x}\mapsto e^{\lambda}\dot{x}$, known in the Finsler geometry literature as the Liouville vector field. Let us focus on the differential form part of the volume form
\begin{align}\label{eq:differentials}
    &i_{\mathbb{C}}\left(\dd \dot \t \wedge \dd \dot r \wedge \dd \dot \theta \wedge \dd \dot \phi\right)\\
    &= \dot \t (\dd \dot r \wedge \dd \dot \theta \wedge \dd \dot \phi) 
    - \dot r (\dd \dot \t  \wedge \dd \dot \theta \wedge \dd \dot \phi)
    + \dot \theta (\dd \dot \t \wedge \dd \dot r \wedge \dd \dot \phi)
    - \dot \phi (\dd \dot \t \wedge \dd \dot r \wedge \dd \dot \theta)\,.
\end{align}
Since we are working on homogeneous and isotropic Finsler spacetimes, it is convenient to employ the following coordinate change
\begin{align}\label{eq:dotr}
    \dot r^{\pm} = \pm \sqrt{(1- kr^2)(s^2 \dot \eta^2 - r^2(\dot \theta^2 + \sin^2\theta \dot \phi^2))}\,.
\end{align}
Moreover, on $\mathcal{S}_x$ we have that $L=1$ and thus, by the assumed conformal separated variable form of the Finsler Lagrangian \eqref{eq:LHomIso}, we have
\begin{align}\label{eq:dt}
    \dot \eta = \pm \frac{1}{a f} \Rightarrow \dd \dot \eta = \mp \left(\frac{\dot a}{a^2 f} \dd \eta + \frac{f'}{a f^2} \dd s\right)\,.
\end{align}
This coordinate transformation is well defined for $s\neq s_I$, where $s_I,\ I=1,2,3,...$ are the roots of $L$
\begin{align}
    L(\t,\dot \t, s_I) = \dot \t^2 a(\t)^2 f(s_I)^2 = 0 \Leftrightarrow f(s_I) = 0\,,
\end{align}
hence, in particular, it is well defined on $\mathcal{S}_x$.

Since we are interested in the integral over $\mathcal{S}_x$ at a fixed spacetime point $x$, we employ that $\dd\t = \dd r = \dd\theta = \dd\phi =0$. Moreover, we focus on the integral over the future pointing unit-timelike directions, thus $\dot \t>0$ on $\mathcal{S}_x$. Therefore, depending on which sign of $\dot r$ in \eqref{eq:dotr} is chosen, the volume form on $\mathcal{S}_x$ is given by the following two expressions,
\begin{align}
    d\Sigma_x^\pm 
    &= \frac{-\det g}{ a(\t)^4 f^4 }\ \ \frac{i_{\mathbb{C}}\left(\dd \dot \t \wedge \dd \dot r \wedge \dd \dot \theta \wedge \dd \dot \phi\right)}{\dot \t^4}\\
    &= \pm \frac{-\det g}{ a^3 f^3 }\ \ \frac{\sqrt{1-kr^2} s}{\sqrt{\frac{s^2}{f^2 a^2}-r^2(\dot \theta^2 + \sin^2\theta \dot \phi^2)}} \dd s \wedge \dd \dot \theta \wedge \dd \dot \phi\,.
\end{align}
Finally, the integral \eqref{eq:int} becomes a sum of two parts, namely integration over $\mathcal{S}_x^+=\mathcal{S}_x|_{\dot r>0}$ and $\mathcal{S}_x^-=\mathcal{S}_x|_{\dot r<0}$, and can be expressed as
\begin{align}\label{eq:int2}
    &\int_{\mathcal{S}_x} \left(\frac{\dot x^{a_1}...\dot x^{a_n}}{L^{n/2}} \varphi(\t,s) \dd\Sigma_x\right)\Big|_{\mathcal{S}_x}\nonumber\\
    &= \int_{\mathcal{S}^+_x} \left(\frac{\dot x^{a_1}...\dot x^{a_n}}{L^{n/2}} \varphi(\t,s) \dd\Sigma^+_x\right)\Big|_{\mathcal{S}^+_x} 
    + \int_{\mathcal{S}^-_x} \left(\frac{\dot x^{a_1}...\dot x^{a_n}}{L^{n/2}} \varphi(\t,s) \dd\Sigma^-_x\right)\Big|_{\mathcal{S}^-_x}\nonumber\\
    &= \int_{\mathcal{S}^+_x}  \left( (\dot x^{a_1}...\dot x^{a_n})|_{\mathcal{S}^+_x}\  \varphi(\t,s) \frac{-\det g}{ a^3 f^3 } \frac{ \sqrt{1-kr^2}\ s}{\sqrt{\frac{s^2}{f^2 a^2}-r^2(\dot \theta^2 + \sin^2\theta \dot \phi^2)}} \dd s \wedge \dd \dot \theta \wedge \dd \dot \phi\right)\nonumber\\
    &- \int_{\mathcal{S}^-_x} \left( (\dot x^{a_1}...\dot x^{a_n})|_{\mathcal{S}^-_x}\ \varphi(\t,s) \frac{-\det g}{ a^3 f^3 } \frac{ \sqrt{1-kr^2}\ s}{\sqrt{\frac{s^2}{f^2 a^2}-r^2(\dot \theta^2 + \sin^2\theta \dot \phi^2)}} \dd s \wedge \dd \dot \theta \wedge \dd \dot \phi\right)\nonumber\\
    &= \int_{\mathcal{S}^+_x}  \left( (\dot x^{a_1}...\dot x^{a_n})|_{\mathcal{S}^+_x}\  \varphi(\t,s) \frac{-\det g}{ a^3 f^3 } \frac{ \sqrt{1-kr^2}\ s}{\sqrt{\frac{s^2}{f^2 a^2}-r^2(\dot \theta^2 + \sin^2\theta \dot \phi^2)}} \dd s \wedge \dd \dot \theta \wedge \dd \dot \phi\right)\nonumber\\
    &+ \int_{\mathcal{S}^+_x} \left( (\dot x^{a_1}...\dot x^{a_n})|_{\mathcal{S}^-_x}\ \varphi(\t,s) \frac{-\det g}{ a^3 f^3 } \frac{ \sqrt{1-kr^2}\ s}{\sqrt{\frac{s^2}{f^2 a^2}-r^2(\dot \theta^2 + \sin^2\theta \dot \phi^2)}} \dd s \wedge \dd \dot \theta \wedge \dd \dot \phi\right)\nonumber\\
    &= \int_{\mathcal{S}^+_x}  \left( \left( (\dot x^{a_1}...\dot x^{a_n})|_{\mathcal{S}^+_x}+(\dot x^{a_1}...\dot x^{a_n})|_{\mathcal{S}^-_x}\right)\  \varphi(\t,s) \frac{-\det g}{ a^3 f^3 } \frac{ \sqrt{1-kr^2}\ s}{\sqrt{\frac{s^2}{f^2 a^2}-r^2(\dot \theta^2 + \sin^2\theta \dot \phi^2)}} \dd s \wedge \dd \dot \theta \wedge \dd \dot \phi\right)\,.
\end{align}

In local coordinates $(\t,r,\theta,\phi)$, we can express the 4-velocity factor in \eqref{eq:int2} by using \eqref{eq:dotr} and \eqref{eq:dt} as
\begin{align}
    (\dot x^{a_1}...\dot x^{a_n})|_{\mathcal{S}^\pm_x} 
    &= (\dot{\eta}^I \dot r^J \dot \theta^K \dot \phi^L )|_{\mathcal{S}^\pm_x}\\
    &=  (a f)^{-I} (\pm1)^J(1- kr^2)^{\frac{J}{2}}\left(\tfrac{s^2}{a^2 f^2} - r^2(\dot \theta^2 + \sin^2\theta \dot \phi^2)\right)^{\frac{J}{2}} \dot \theta^K \dot \phi^L \,,
\end{align}
with $I + J + K + L = n>0$, $I,J,K,L \in \mathbb{N}$. For the integration, we are interested in the sum of the terms on $\mathcal{S}^+_x$ and $\mathcal{S}^-_x$, which becomes
\begin{align}\label{eq:dotxn}
    &(\dot x^{a_1}...\dot x^{a_n})|_{\mathcal{S}^+_x} + (\dot x^{a_1}...\dot x^{a_n})|_{\mathcal{S}^-_x}\nonumber\\
    &= (a f)^{-I} (1 + (-1)^J)(1- kr^2)^{\frac{J}{2}}\left(\tfrac{s^2}{a^2 f^2} - r^2(\dot \theta^2 + \sin^2\theta \dot \phi^2)\right)^{\frac{J}{2}} \dot \theta^K \dot \phi^L\,.
\end{align}
In order to further evaluate the integrals \eqref{eq:int2} it is convenient to introduce new coordinates $\dot v$ and $\dot \alpha$ for $\dot \theta$ and $\dot \phi$ by
\begin{align}
    r \dot \theta  = \dot v \cos \dot \alpha\,,\quad 
    r \sin\theta \dot \phi  = \dot v \sin \dot \alpha\,,\quad
    \frac{s}{f a}>\dot v>0\,, \dot \alpha \in (0,2\pi)\,.
\end{align}
They lead to 
\begin{align}
   r^2(\dot \theta^2 + \sin^2\theta \dot \phi^2) = \dot v^2\,,\quad
   \dd \dot \theta \wedge \dd \dot \phi = \dot v \dd \dot v \wedge \dd \dot \alpha\,.
\end{align}
Moreover, $s\in (s_1,s_2)$, where $s_1<0$ and $0<s_2$ are the roots of $L$ closest to $0$.

Using \eqref{eq:dotxn} and the new coordinates, the integral \eqref{eq:int2} can be written as
\begin{align}\label{eq:momentIntGen}
        &\int_{\mathcal{S}_x} \left(\frac{\dot x^{a_1}...\dot x^{a_n}}{L^{n/2}} \varphi(\t,s) \dd\Sigma_x\right)\Big|_{\mathcal{S}_x}\nonumber\\
        &= \int_{0}^{s_2} \int_0^{\frac{s}{f a}} \int_0^{2\pi} \mathcal{I}(\t,r, s, \dot v, \dot \alpha)\ \varphi(\t,s)\ \dd s \wedge \dd \dot v \wedge \dd \dot \alpha\,,
\end{align}
with
\begin{align}\label{eq:momentIntGen2}
    \mathcal{I}(\t,r, s, \dot v, \dot \alpha) 
    &= - s \det g   \nonumber \\
    &\times \left(  \frac{(1 + (-1)^J)}{(a f)^{I+3}} (1- kr^2)^{\frac{J+1}{2}}\left(\tfrac{s^2}{a^2 f^2} - \dot v^2\right)^{\frac{J-1}{2}} \frac{\dot v^{K+L+1}}{r^{K+L}} \cos^K(\dot \alpha) \sin^L(\dot \alpha)\right)\,.
\end{align}
The $\dot \alpha$ and $\dot v$ integration in \eqref{eq:momentIntGen} can be carried out in terms of Euler $\Gamma$ functions 
\begin{align}\label{eq:momentIntGen3}
    &\int_{\mathcal{S}_x} \left(\frac{\dot x^{a_1}...\dot x^{a_n}}{L^{n/2}} \varphi(\t,s) \dd\Sigma_x\right)\Big|_{\mathcal{S}_x}\nonumber\\
    &= \left(1 + (-1)^J\right) \left(1+(-1)^K\right)\left(1+(-1)^{K+L}\right) \frac{\Gamma\left(\tfrac{(1+K)}{2}\right)\Gamma\left(\tfrac{(1+L)}{2}\right)\Gamma\left(\tfrac{(1+J)}{2}\right)}{4 \Gamma\left(\frac{(3+J+K+L)}{2}\right)}\nonumber\\
    &\times\int_{0}^{s_2}  \mathcal{J}(\t,r, s)\ \varphi(\t,s)\ \dd s \,,
\end{align}
with
\begin{align}\label{eq:momentIntGen3b}
    \mathcal{J}(\t,r, s)   
    &=  \det(w_{\alpha\beta})  \frac{(1- kr^2)^{\frac{J+1}{2}} }{r^{K+L}} a^{8} f^{5} f'^2 (-f'')  
    \frac{s^{J+K+L}}{(a f)^{4+I+J+K+L}} \,.
\end{align}

The expressions \eqref{eq:momentIntGen3} and \eqref{eq:momentIntGen3b} are the key to evaluate the different moment integrals \eqref{eq:int} by fixing the exponent $I$ of $\dot \eta$, $J$ of $\dot r$, $K$ or $\dot \theta$ and $L$ or $\dot \phi$.

\subsection{The scalar integral}\label{sapp:scalint}
For the scalar integral, which can be used to evaluate \eqref{eq:NumbDense}, we set $n=0$  in \eqref{eq:int2} or $I=J=K=L=0$ in \eqref{eq:momentIntGen3} and \eqref{eq:momentIntGen3b}, to find
\begin{align}\label{eq:int3}
    \hat N &= \int_{\mathcal{S}_x} \left(\varphi(\t,s) \dd\Sigma_x\right)|_{\mathcal{S}_x}\nonumber\\
    &=4\pi  a^{4}\sqrt{1-kr^2} \det(w_{\alpha\beta})
    \int_{s_0}^{s_1} \dd s \left( f  f'^2 (-f'')\ \ \varphi(\t,s) \right)\,.
\end{align}

\subsection{The vector integral}\label{sapp:vecint}
Setting $n=1$ in \eqref{eq:int2} or one of the exponents $I$, $J$, $K$ or $L$ to $1$ in \eqref{eq:momentIntGen3} and \eqref{eq:momentIntGen3b}, we find the four first moments \eqref{eq:ParticleCurrentDense}:
\begin{align}\label{eq:intVec1}
    \hat A^0 &= \int_{\mathcal{S}_x} \left(\dot \t \varphi(\t,s) \dd\Sigma_x\right)\Big|_{\mathcal{S}_x}\nonumber\\
    &=4\pi a^{3} \sqrt{1-kr^2} \det w_{\alpha\beta} \int_{s_0}^{s_1} \dd s \left(  \varphi(\t,s)  f'^2 (-f'')  \right)\,.
\end{align}
For the first moments with either $J=1$, $K=1$, or $L=1$ and all other exponents being zero, it is clear from \eqref{eq:momentIntGen3} that they all vanish
\begin{align}\label{eq:intVec2}
    \hat A^1 &= \int_{\mathcal{S}_x} \left(\dot r \varphi(\t,s) \dd\Sigma_x\right)\Big|_{\mathcal{S}_x} =  0\,,\\
    \hat A^2 &= \int_{\mathcal{S}_x} \left(\dot \theta \varphi(\t,s) \dd\Sigma_x\right)\Big|_{\mathcal{S}_x} = 0\,,\\
    \hat A^3 &= \int_{\mathcal{S}_x} \left(\dot \phi \varphi(\t,s) \dd\Sigma_x\right)\Big|_{\mathcal{S}_x} = 0\,.
\end{align}

The higher moment integrals can be discussed in an analogue way.

\bibliographystyle{JHEP}
\bibliography{FFSeparated}

\end{document}